\documentclass[lettersize,journal]{IEEEtran}
\usepackage{amsmath,amsfonts}
\usepackage{algorithmic}
\usepackage{algorithm}
\usepackage{algorithmic}
\usepackage{array}
\usepackage[caption=false,font=normalsize,labelfont=sf,textfont=sf]{subfig}
\usepackage{textcomp}
\usepackage{stfloats}
\usepackage{url}
\usepackage{verbatim}
\usepackage{graphicx}
\usepackage{xcolor}

\usepackage{amsthm}  

\newtheorem{definition}{Definition} 
\newtheorem{theorem}{Theorem}
\newtheorem{lemma}[theorem]{Lemma}
\newtheorem{example}{Example}
\usepackage{multirow}

\usepackage{adjustbox}

\def\BibTeX{{\rm B\kern-.05em{\sc i\kern-.025em b}\kern-.08em
    T\kern-.1667em\lower.7ex\hbox{E}\kern-.125emX}}
\usepackage{balance}
\begin{document}
\title{Decentralized Pliable Index Coding For Federated Learning In Intelligent Transportation Systems}
\author{Sadina Kadakkottiri,~\IEEEmembership{Member,~IEEE}, Narisetty Harish, Nujoom Sageer Karat, \IEEEmembership{Member,~IEEE}, Deepthi Paramel  Pattathil,~\IEEEmembership{Senior Member,~IEEE}, Balaji Sundar Rajan,~\IEEEmembership{Fellow,~IEEE}


\thanks{Sadina Kadakkottiri, Deepthi Paramel Pattathil and Nujoom Sageer Karat are with ECE Department, National Institute of Technology, Calicut, 673601, India (e-mail: sadinasideek@gmail.com, nujoom@nitc.ac.in, deepthi@nitc.ac.in).}
\thanks{Narisetty Harish is currently working in Synopsis, Inc. The work was done when he was at NIT Calicut. (e-mail:  nharish000@gmail.com ).}
\thanks{ Balaji Sundar Rajan is with the Department of Electrical Communication Engineering, Indian Institute of Science, Bangalore, 560012, India (e-mail:  bsrajan@iisc.ac.in).}}


\maketitle

\begin{abstract}
Federated Learning (FL) is becoming increasingly popular in many application areas, including smart city monitoring, autonomous driving, anomaly detection, and disaster management.
FL is a promising option for data privacy and security in Intelligent Transportation Systems (ITS), because it allows edge devices, Road Side Units (RSUs), and Central Server (CS) to jointly train the machine learning model. Since RSU collects data from the vehicles passing through its range, the local data of each RSU will have a non-IID distribution, which adversely affects the convergence speed and accuracy of FL training as discussed in the work by Li \textit{et al.} (``FEEL: Federated End-to-End Learning With Non-IID Data for Vehicular Ad Hoc Networks,'' \textit{IEEE Transactions on Intelligent Transportation Systems}, 2022). Generating synthetic data locally at
individual nodes, followed by data shuffling among the nodes, is a promising approach to address the Non-IID data problem. In this work, we propose pliable index coding (PIC) solutions for efficient data shuffling among the nodes in an FL system. In PIC($S$) problems, a client is satisfied if it can retrieve any $S$ new
messages not originally present in its side-information. We particularly consider decentralized pliable index coding problems (DPIC) where the clients communicate among themselves without a central server to model the data shuffling in FL. A class of DPIC, known as Consecutive Decentralized Pliable Index Coding (CDPIC($S$,$K$)), where each client has $K$ consecutive messages as side-information, is considered. For CDPIC($S$,$K$) problems, pliable index code designs are provided for any value of $K$ and $S$, and optimality proofs for some of the cases are established. Further, these CDPIC solutions are applied for data shuffling in FL, to transform the local data distribution towards IID progressively with each transmission, thereby enhancing the performance of FL. The improvement in the accuracy and convergence of the most popular FL technique, FedAvg, and a promising federated submodel technique, CELL (Communication Efficient Lottery Learning), are analysed by providing different degrees of data shuffling using the proposed CDPIC schemes.


\end{abstract}


\section{Introduction}
\IEEEPARstart{A}{dvancements} in 5G technology offer significant improvements in processing and sensing for vehicular networks (VN) \cite{its}. Recent research in Intelligent Transportation Systems (ITS) has increasingly focused on advancements in vehicular networks, incorporating emerging technologies like 5G, the Internet of Things (IoT), edge computing, and cloud computing \cite{securyneed}. Key components of ITS include vehicle nodes, sensors,  Road Side Units (RSUs), and main base stations \cite{FLinedge}. As vehicles are equipped with various sensors, actuators, GPS devices, and onboard computers, they can communicate with both mobile and stationary RSUs, which possess significant computational power \cite{vehicular2014}. The integration of connected vehicular networks and automated vehicles to address critical challenges like traffic accidents, congestion, and pollution is a rapidly advancing domain \cite{surveyitsautamatedvehicles}.

 Recently, Machine Learning (ML) applications in ITS have been widely researched.A detailed survey of Deep Learning (DL) applications in autonomous driving is presented in  \cite{DLautonomous}. ML approaches in VNs involve collecting large volumes of data and processing them at a central node to address ITS challenges. While effective, this method introduced substantial communication and computational overheads \cite{centralizedtoFL}. The development of Distributed Machine Learning (DML) emerged as a solution, leveraging the computational capabilities of edge devices \cite{hu2021distributed}. However, DML raised concerns related to privacy and communication overhead \cite{FLinITSopenprob}. To address these issues, Federated Vehicular Networks (FVNs) were proposed, utilizing stable vehicle connections and Dedicated Short-Range Communications (DSRC) to support DML \cite{FLopertunities}. FL further mitigates the challenges of DML by allowing edge devices to collaboratively train models without sharing raw data, thus preserving privacy \cite{fed23}. One of the most commonly used algorithms in FL is FedAvg, where a global model is distributed to worker nodes for local training, and the aggregated model parameters are updated centrally in an iterative manner \cite{mcmahan2017communication}.

However, FVNs face significant challenges, such as data quality degradation due to the dynamic nature of networks, the mobility of participating nodes, and their varying capabilities \cite{FLinvehicularappli}. In many ITS applications, data collected by edge devices (such as vehicles or RSUs) do not follow an independent and identically distributed (IID) pattern, leading to suboptimal local training performance \cite{ITSnoniid}. For instance, in autonomous driving scenarios, vehicles may collect data from varied environments, such as urban areas, highways, and residential roads, making it difficult to assume IID data. This non-IID nature of data distribution can result in degradation of convergence performance, causing reduced accuracy and increased latency \cite{mrad2021federated}. Thus, in a delay-sensitive ITS scenario, it becomes challenging to meet the performance demands of FL. To address this, several solutions have been proposed. For instance, an FL algorithm combined with reinforcement learning was introduced to mitigate these challenges \cite{fedheteroge}.
Methods like FedProx \cite{li2020federated}  as well as asynchronous federated learning optimization techniques \cite{xie2019asynchronous} have been explored to improve performance when the data distribution is heterogeneous. Federated Submodel Learning (FSL) techniques such as FedRolex \cite{fedrolex}, CELL \cite{CELL} are also developed to address the challenges of heterogeneous data distribution. In FSL, each client extracts a sub-model from the global model and locally trains the sub-model with available data. Extraction of sub-models with varying capacities suitable for local data distribution from the large global model helps in reducing transmission cost and improving the speed of convergence.

\begin{figure*}[!t]
\centering
\includegraphics[width=8in]{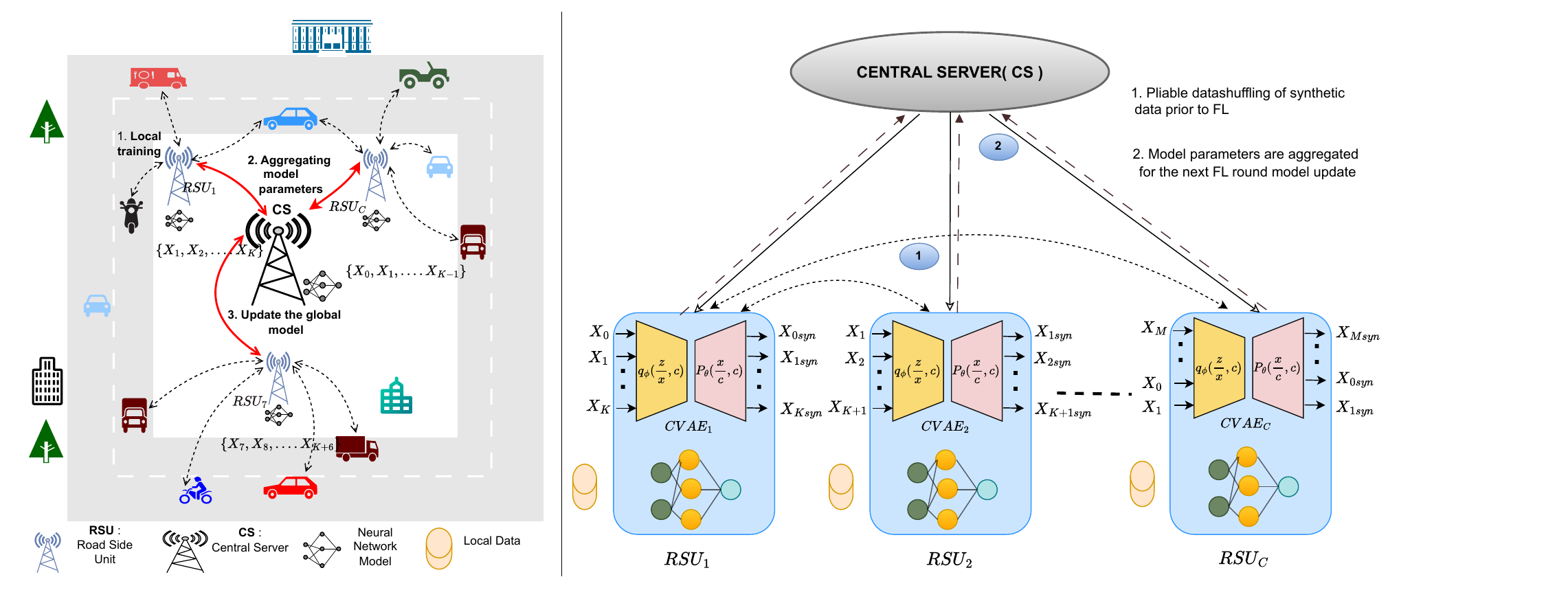}
\caption{Framework of FL in vehicular networks: Consider a rectangular/ circular region around the CS; $C$ RSUs act as nodes, each RSU senses local data from vehicles passing through its range, giving rise to non-IID data distribution among RSUs; each RSU constitutes a CVAE to generate synthetic data corresponding to their local data, synthetic data of each data class is transmitted to other nodes using proposed pliable data shuffling methods. Then, each RSU will perform local training with its own sensed data and received synthetic data.}
\label{fig_systm}
\end{figure*}
\vspace{-0.18em}

When the data distribution is highly non-IID in the sense that individual clients have access to data samples from only a few classes, these approaches demand many rounds of communication, which results in increased latency. Another popular research direction to reduce the delay without affecting the accuracy and privacy constraints is to perform data shuffling on synthetic data. This generation of synthetic data plays a vital role in this VANET scenario. 
 Various generative models, including Generative Adversarial Networks (GANs) and Variational Autoencoders (VAEs), have been widely explored for this purpose. Among them, Conditional GAN (CGAN), Wasserstein GAN (WGAN), Conditional Wasserstein GAN (CWGAN), and Conditional Variational Autoencoder (CVAE) have demonstrated promising capabilities in generating high-quality synthetic data \cite{ganvae}. While VAEs effectively capture latent representations to generate diverse samples \cite{vae}, GANs employ a generator-discriminator architecture to synthesize data that closely mimics real distributions \cite{gan}. In \cite{li2024synthetic}, authors have shown that data shuffling with synthetic data can be used to improve performance in FL when data distribution is heterogeneous among nodes. But they have used diffusion models for synthetic data generation. Diffusion models have high computational complexity, which makes them unsuitable for delay-sensitive applications and edge computing.

To identify the most suitable approach for privacy-preserving synthetic data generation, we conducted a comparative analysis of these models based on key criteria, including model architecture, privacy mechanisms, algorithm complexity, training requirements, and performance in our earlier work \cite{snehasri}. In that work, we focused on 
 developing a generative model that ensures high-quality synthetic data generation while maintaining minimal structural complexity, making it feasible for deployment on edge devices in vehicular networks. It was observed that CVAE-based synthetic data generation offers low structural complexity and sufficient data quality.
Therefore, RSUs can effectively leverage CVAE-based models for generating synthetic data corresponding to the local data of each node. This synthetic data is utilized in the data shuffling phase of the work discussed in this paper to ensure privacy and diversity among local datasets. Moreover, we develop pliable index coding-based solutions for low-latency data shuffling. By leveraging efficient data shuffling, our approach not only improves the target accuracy of FL but also significantly reduces the number of communication rounds required to meet this target accuracy.

The remainder of the paper is organized as follows. Section \ref{sec:prelim} discusses the preliminaries of the pliable index coding and FL. The system model considered and the major contributions of this research work are detailed in Section \ref{sec:system_model}. Section \ref{sec:main} discusses the main results related to decentralized pliable index coding, and the corresponding code constructions are discussed in Section \ref{sec:code}. The application of these codes to the FL scenario and the corresponding results are illustrated in Section \ref{sec:fedavg_appl}. Finally, we conclude the paper in Section \ref{sec:conclusion}.

\section{Preliminaries}
\label{sec:prelim}
In this section, we discuss the preliminaries of decentralized pliable index coding, federated learning, federated submodel techniques and generative
methods.  

\subsection{Decentralized Pliable Index Coding}

Index Coding with side information \cite{bar2011index} is a problem in communication networks involving a central server (CS) and multiple clients, each having access to certain side information. The goal is to minimize the number of transmissions required to meet all client demands. In this problem, a server broadcasts linear combinations of $M$ messages $\{X_0, X_1, \ldots, X_{M-1}\}$ over a noiseless channel to $C$ clients, where each client has a subset of the messages as side information and requests certain messages it does not possess. The objective is to devise a coding strategy that meets all client demands with the smallest number of possible transmissions. The index coding problem is known to be NP-hard \cite{PeetersNP, DauNP}, with solutions available only for specific cases.

In some practical applications, such as internet search, client demands are flexible—any new message that a client lacks may satisfy its need \cite{Brahmapliable}. This variant, known as the pliable index coding problem, is also NP-hard but offers exponential savings over conventional index coding \cite{Brahmapliable}. Various works in the literature present code constructions for pliable index coding \cite{Songpliable, Fu_pliable, Krishnan_pliable, Eghbal_pliable}. Pliable index coding can reduce communication load in data shuffling for distributed computing systems \cite{Songpiabledata}. The paper \cite{sasi2019code} explores code constructions for the pliable index coding problem with consecutive side information, detailing approaches for cases where each client receives one desired message, and where the total number of decoded messages across clients is maximized.

Decentralized Pliable Index Coding (DPIC), introduced in \cite{decpliable}, extends this concept to scenarios where clients communicate among themselves without a central server. DPIC finds wide applications in peer-to-peer content sharing, but only a few studies discuss code construction for DPIC \cite{decpliable, Liu_secure_dPIC, Liu_secure_code}. In this work, we consider the application of a special class of decentralized pliable index coding problem, namely CDPIC($S$,$K$), to improve FL convergence in vehicular networks. In CDPIC($S$,$K$), each client has $K$ consecutive messages as side information, and its data demands will be satisfied upon receiving $S$ new messages.

The following definitions clarify key terminologies:

\begin{definition}
A DPIC($S$) problem consists of $M$ messages and $C$ clients without a central server. Each client holds a subset of messages as side information and is satisfied upon receiving $S$ new messages that it did not previously possess.
\end{definition}

\begin{definition}
A CDPIC($S$,$K$) problem consists of $M$ messages and $C$ clients without a central server. Each client possesses a set of $K$ consecutive messages as side information. Specifically, the $i$-th client has the side information set $\{X_i, X_{i+1}, \ldots, X_{i+K-1}\}$. Each client is satisfied upon receiving $S$ new messages.
\end{definition}

\subsection{Federated Learning}
McMahan \textit{et al.} \cite{mcmahan2017communication} introduced FedAvg, a federated optimization method where local data training occurs on a distributed network of edge devices, and a shared model is created by combining locally calculated updates. Federated optimization faces many challenges compared to traditional distributed optimization due to issues in data distributions such as non-IID data, unbalanced data and broadly dispersed data. 
Several studies \cite{ITSnoniid}\cite{mrad2021federated} show that the class imbalance and non-IID nature of data 
 adversely affect FL performance in terms of convergence speed and accuracy. So, in \cite{li2024synthetic}, it was demonstrated that shuffling a fraction of locally generated synthetic data across clients can significantly reduce the gradient dissimilarity and accelerate convergence, thus improving FL performance. Data shuffling demands additional communication rounds, thereby causing increased latency. However, it was shown that index coding techniques can be efficiently utilized to minimize the delay in data shuffling among edge devices\cite{lee2017speeding} \cite{song2019pliable}. 

\subsection{Federated Submodel Techniques}
In conventional FL, worker nodes receive the latest global model and update it using their local data in each communication round. These locally trained models are then transmitted back to a central server, where updates are aggregated to refine the global model for the next round. However, as the number of communication rounds increases, especially with a large client base and deep learning tasks, transmission overhead becomes a significant concern, particularly in resource-constrained networks such as VANETs and IoT systems.

To address this challenge, federated submodel learning was introduced, where clients download only the necessary parts of the global model (submodels) instead of the full model, thereby reducing communication costs \cite{niu2020billion}. Several submodel extraction methods are developed based on partial training (PT), which rely on random or static selection strategies for submodels. However, this can lead to client drift, where differences between client model architectures and the global model result in unevenly trained parameters. The lottery ticket hypothesis suggests that within a neural network, there exist smaller, well-optimized subnetworks (lottery tickets) that can achieve similar accuracy as the full model while reducing computational and storage requirements. LotteryFL leverages this idea by assigning subnetworks of the global model to each client. However, it relies on unicasting, which prioritizes fast learners while neglecting straggler nodes, leading to inefficiencies in bandwidth-constrained environments \cite{lotteryfl}.

To overcome these limitations, CELL (Communication Efficient Lottery Learning) introduces an improved pruned FL approach that optimizes transmission overhead while maintaining accuracy. Instead of unicasting different subnetworks, CELL broadcasts the full global model to all clients, allowing each client to selectively prune subnetworks based on their validation accuracy. Clients with high accuracy thresholds refine their subnetworks through lottery learning, while others continue training the full model, ensuring collective improvement over time \cite{CELL}. These optimized subnetworks are then aggregated, reducing the number of parameters transmitted to the server. Moreover, CELL’s downlink broadcasting strategy significantly reduces communication costs while achieving FL-level accuracy, making it well-suited for large-scale and resource-limited FL environments.
 Thus, CELL helps to improve the speed of
convergence and total power
consumption, especially in the context of diverse and dynamic vehicular environments.

\subsection{Generative Methods}
 Synthetic data generative units are incorporated into each node of FL to create synthetic data that maintains the statistical properties of the original data, making it suitable for tasks like data shuffling in FL without exposing sensitive information \cite{li2024synthetic}. GANs consist of a generator, which creates synthetic samples, and a discriminator, which distinguishes between real and
 generated samples. Variants like CGANs and WGANs \cite{gananalysis} enable
 improved training stability and conditional data generation. A generator consists of  3-5 fully
 connected layers, each followed by batch normalization and
 Leaky ReLU activations, ending with a tanh layer for realistic
 data generation. The discriminator has fully connected
 layers with Leaky ReLU activations and a final sigmoid layer
 to classify real versus generated data. 
  VAEs use a probabilistic approach to map input data into
 a latent space and generate new samples by sampling from
 this space, enabling effective data reconstruction with privacy
 protection \cite{vae}.  The encoder consists of 2-4 convolutional layers, capturing
 data features in a latent space while utilizing batch normalization and ReLU activations. The decoder mirrors the encoder’s
 architecture, employing deconvolutional layers to reconstruct
 data from the latent representation.
  On comparing the architectural differences of GAN variants and CVAE, it is seen that CVAE is more suitable for data generation in edge devices due to the lesser complexity in their architectures and reduced latency \cite{snehasri}.

\section{System Model and Major Contributions}
\label{sec:system_model}
The system model in this work considers an ITS setting similar to that discussed in \cite{enhancedFL}. Consider a FL system for applications such as road detection in autonomous driving \cite{surveyitsautamatedvehicles} and smart city integration \cite{al2023federated}. The roads in a rectangular/circular region around a central server (CS), as given in Fig. \ref{fig_systm}, are considered with RSUs as worker nodes sensing data through vehicles as edge devices.
 Each RSU will collect data from multiple vehicles in its range. Now the road covered in the surveillance area of the CS includes different geographies such as city area, highway, suburban area, and residential area, giving rise to data samples from the corresponding classes. Each RSU  will collect the data samples corresponding to its coverage area. Since the data collected by each RSU will correspond to only a few of these different classes, the data samples will have a non-IID distribution. Here, it is assumed that the surveillance areas of consecutive RSUs overlap to ensure effective learning for autonomous driving, smart city integration, or other similar applications. Based on the coverage area of an RSU, the number of data classes available with an RSU changes; depending on the positioning of the RSUs, the amount of overlap between two consecutive RSUs will vary.  Now, it is required to carry out a data shuffling among RSUs to improve the performance of FL. The amount of data that needs to be redistributed among the RSUs depends on the number of data classes at an RSU and the class overlap among consecutive RSUs. 
 
 Consider an ITS system consisting of $C$ RSUs that collect data from $M$ classes, where each RSU collects local data from $K$ consecutive classes. Let $M$ classes of messages be denoted by $\{X_0, X_1, \ldots, X_{M-1}\}$ and $C$ clients denoted by $\{C_0, C_1, \ldots, C_{C-1}\}$. The first group of edge vehicles connected to RSU$_0$ collects $K$ consecutive classes of data namely,  ${X_0, X_1, \ldots, X_{K-1}}$, and sends it to RSU$_0$ as ${D_0}$. The next RSU collects a subsequent set of consecutive data, ${X_1, \ldots, X_{K}}$, as ${D_1}$, and this pattern continues. In this way, each RSU$_i$ holds a local dataset $D_i$. Here, $K$ is determined by the range of the RSU. When $ K < M$, the data is referred
to as non-IID, resulting in slow convergence and low accuracy in FL. The amount of data samples each RSU requires to make its data distribution close to IID is indicated by $S$.  This then corresponds to the CDPIC($S$,$K$) problem with $M$ messages and $C$ clients,
 where each client has $K$ consecutive messages and each client is satisfied on receiving $S$ new messages that they don't have.

\subsection{Proposed FL Algorithm with Synthetic Data Generation and CDPIC-Based Data Shuffling}

We propose an efficient FL approach that enhances performance using synthetic data shuffling while minimizing communication overhead. The ITS setup considered in this work assumes non-IID data distribution among the RSUs, which corresponds to the CDPIC($S$,$K$) problem. The data required for shuffling is generated synthetically at each node. Therefore, in our system model, each RSU is equipped with a synthetic data generation unit. After collecting local data (before local training), each RSU generates synthetic data, which is then used for shuffling with proposed CDPIC($S$,$K$) algorithms. 

The algorithm for the proposed FL with synthetic data generation and pliable data shuffling is given in Algorithm \ref{alg:alg1}. The algorithm contains 6 phases, out of which the initial 4 phases are considered as pre-processing stages that need to be completed before applying the proposed CDPIC($S$,$K$) algorithms for data shuffling. The Phase 1 deals with the choice of network architecture and the FL algorithm for the application. The network architecture should be chosen considering the complexity of the dataset and the communication overhead of the FL process. This work considers two network architectures, say CNN and lightweight CNN. Lightweight CNN has a much smaller number of parameters compared to CNN, thereby helping to reduce the communication overhead of the FL process significantly. But for complex datasets such as CIFAR10, a lightweight CNN may not be able to provide the required levels of accuracy. In the same way, the choice between FL and FSL algorithms needs to be made considering the tolerable delay and communication overhead. While the FL process requires a larger number of parameters to be transmitted between RSUs and CS in each communication round, the number of communication rounds is higher for FSL.  This work considers FedAvg \cite{mcmahan2017communication} as FL algorithm  and CELL \cite{CELL} as FSL algorithm and compares the performance. Phase 2 deals with the data augmentation. In an ITS set-up, different RSUs will get data from vehicles that pass through its range. Hence, there could be an imbalance in the number of data samples available/class across various nodes. For improved FL performance, this data imbalance needs to be addressed through suitable data augmentation methods. Further to this, the number of  synthetic data samples that need to be generated/class at each node is assessed in Phase 3 and the synthetic data generation is carried out in Phase 4. To decide the generative model to be used in this process, we conducted an initial research \cite{snehasri}.  Based on this research, we use CVAE as the generative model  in this work to ensure computational efficiency,  making it suitable for resource-limited nodes. In Phase 5, data shuffling is carried out by using proposed CDPIC($S$,$K$) algorithm. This approach ensures efficient synthetic data shuffling while addressing data heterogeneity in the system. Finally, Phase 6 runs the chosen FL algorithm where each RSU performs local training and transmits updates on model parameters. The CS then aggregates the local model parameters from  all RSUs, updates the global model, and this process repeats until the training converges.


\begin{algorithm}[htbp]
\small
\caption{Proposed Federated Learning Algorithm With CDPIC-based Synthetic Data Shuffling} 
\label{alg:alg1}
\begin{algorithmic}
\STATE \textit{\textbf{Input:} $C$= Number of nodes, $M$= Total number of data classes, $K$= Number of data classes possessed by each node, $P_{ij}$ = Number of data samples in $i^{th}$ class at $j^{th}$ node, $P$ =Number of samples/class required for IID performance for FL with C nodes, $S$= Number of required data classes to get CML accuracy }
\STATE \textit{\textbf{Output:} Trained global model.}

\STATE {\textsc{Pre-processing Stages:} PHASE 1 to 4}
\STATE {\textsc{Phase 1: Model and Algorithm Selection}}
\STATE \hspace{0.87cm} 1.1 Choose a suitable deep learning model.
\STATE \hspace{0.87cm} 1.2 Based on $C$, $M$, and $K$, and considering resource\\\hspace{1.2cm} constraints and performance requirements, choose a \\\hspace{1.2cm} suitable FL algorithm.

\STATE {\textsc{Phase 2: Data Augmentation}}
\STATE \hspace{0.93cm} 2.1 Simulate the original data distribution with $C$ nodes \\\hspace{1.4cm} and $M$ data classes.
\STATE \hspace{0.93cm} 2.2 If $P_{ij} < P$, perform data augmentation to ensure each\\\hspace{1.4cm} of $K$ available classes contain at least $P$ samples


\STATE {\textsc{Phase 3: Evaluation of data transmission requirement }}
\STATE \hspace{0.93cm} 3.1 Based on the data distribution, model architecture and\\\hspace{1.5cm}the FL algorithm, determine 
 the required number of  \\ \hspace{1.5cm}data classes $S$ to  get the CML accuracy.
\STATE {\textsc{Phase 4: Synthetic Data Generation}}
\STATE \hspace{0.93cm} 4.1 Generate synthetic data samples using a Conditional\\\hspace{1.5cm} Variational Autoencoder (CVAE).
\STATE {\textsc{Phase 5:
Data Shuffling and Transmission Scheme}}
\STATE \hspace{0.9cm} 5.1 \textbf{if} $M = C$ \textbf{then}
\STATE \hspace{0.9cm} 5.1.1 \textbf{if} $K \leq \left\lfloor \frac{M}{S+1} 
\right\rfloor$, \textbf{then}
 \\\hspace{2.2cm} Convey $S$ synthetic data using uncoded \\\hspace{2.3cm}transmission scheme (Sec. \ref{K_less_M2_a}).

\STATE \hspace{0.9cm} 5.1.2 \textbf{elseif} $(n{-}1) \left\lfloor \frac{M}{S{+}n{-}1} \right\rfloor < K \leq n \left\lfloor \frac{M}{S{+}n} \right\rfloor \leq \left\lfloor \frac{M{+}2}{3} \right\rfloor$,\\ 
\hspace{2.3cm} \textbf{where} $n \in \{2, 3, \ldots, M{-}S\}$ , \\ 
\hspace{2.3cm} Convey $S$ synthetic data using uncoded \\ 
\hspace{2.3cm} transmission scheme (Sec. \ref{K_less_M2}).

\STATE \hspace{0.9cm} 5.1.3 \textbf{elseif}
$ \lfloor \frac{M+2}{3} \rfloor < K < \lfloor\frac{M}{2}\rfloor$, \textbf{then} 

\hspace{2.3cm} Convey $S$ synthetic data using coded \\\hspace{2.3cm}transmission scheme (Sec. \ref{K_less_M2_b}).


\STATE \hspace{0.9cm} 5.1.4 \textbf{elseif} $K\in \{\lfloor\frac{M}{2}\rfloor,\lfloor\frac{M}{2}\rfloor+1$\}, \textbf{then}\\
\hspace{2.2cm}Convey $S$ synthetic data using coded \\\hspace{2.3cm}transmission scheme (Sec. \ref{K_eqM2}).
\STATE \hspace{0.9cm} 5.1.5 \textbf{else} $K > \frac{M}{2}+1$ , \textbf{then}\\
\hspace{2.2cm} Convey $S$ synthetic data using coded \\\hspace{2.3cm}transmission scheme(Sec. \ref{K_grM2}).
\STATE \hspace{0.9cm} 5.1 \textbf{end if}
\STATE \hspace{0.9cm} 5.2 \textbf{else} $M > C$, \textbf{then}\\
\hspace{2.3cm} Convey $S$ synthetic data using uncoded/coded \\\hspace{2.3cm} using transmission scheme(Sec. \ref{MgrthC}).



\STATE \hspace{0.9cm} 5.2 \textbf{end if}

\STATE {\textsc{Phase 6: Federated Learning Execution}}
\STATE \hspace{0.9cm} 6.1 Run the chosen FL algorithm.
\end{algorithmic}
\end{algorithm}

\subsection{Major Contributions}
To the best of our knowledge, this is the first work that proposes Pliable Index Coding solutions to improve the performance of FL algorithms.

The major contributions are as follows:
\begin{itemize}
\item {Established a lower bound for the general Decentralized Pliable Index Coding problem DPIC($S$), demonstrating that at least $S+1$ transmissions are required to satisfy client demands.}
   
\item {Developed optimal  index code solutions  for the CDPIC($S$,$K$) problem for different values of  $K$  and major results summarized in Table \ref{tab:mytable}.}

\item {Using the CDPIC($S$,$K$) solutions, developed optimal transmission schemes to pliably shuffle data in the Federated Learning setting of ITS, resulting in a considerable improvement in the convergence of FL in terms of speed and accuracy.}

\item {Presented a comparative analysis of improvement in the convergence performance of FedAvg and CELL by data shuffling using the proposed CDPIC schemes and tabulated the performance improvement in communication efficiency. }

\end{itemize}

\section{Main results on DPIC}
\label{sec:main}
In this section, we summarize the main results obtained in pliable index coding in this work. We consider the decentralized pliable index coding problem and we present the major results.
The first two theorems are for the general DPIC($S$) problem, and the rest of the main results are for CDPIC($S,K$) problem. In these discussions, the minimum achievable number of transmissions required to satisfy the demands of each DPIC client with $S$ new messages is denoted as $N$. The number of clients that get a new message from the $i$-th transmission is denoted as $R_i$. 
\begin{theorem}
\label{thm:lowerbound}
    The number of transmissions for DPIC($S$) should be at least $S+1$.
\end{theorem}

\begin{IEEEproof}
Consider the client who makes the first transmission. It is clear that this client will not receive any new message from the first transmission. It requires $S$ new messages that would come from at least $S$ independent transmissions. Therefore, a total of $S+1$ transmissions are required for a DPIC($S$) problem.
\end{IEEEproof}
\begin{theorem}
\label{thm:basic}
    The total number of clients (repetition allowed) who get a new message from $N$ DPIC transmissions is given by
    $$ \sum_{i=1}^{N} R_i= CS.$$
\end{theorem}

\begin{IEEEproof}
Since each client requires $S$ new messages, the total number of new messages to be received by $C$ clients, so that their demands are satisfied, is $CS$. The $N$ transmissions should facilitate the reception of these $CS$ new messages. If $R_i$ clients decode a new message from the $i$-th transmission, the total number of new messages decoded from $N$ transmissions $\sum_{i=1}^{N} R_i,$ needs to be $CS$.   
\end{IEEEproof}

\begin{theorem}
\label{Thm:uncoded}
    If the number of clients $C$ satisfies $K \leq \frac{C+2}{3},$ transmitting uncoded messages are optimal for CDPIC($S$,$K$).
\end{theorem}
\begin{IEEEproof}
    Since each client has $K$ messages in the side information set, and the messages possessed are consecutive, each message $X_i$ is present in the side information sets of exactly $K$ clients. Hence, if transmitted messages are uncoded, the number of clients $N_U$ served by a transmission, say $X_i$, is at most $C-K$. For transmitted messages to be coded with two messages as $X_i + X_j$, there needs to be at least one node that possesses both the messages $X_i$ and $X_j$.  Then the number of clients possessing only one of the messages $X_i$ or $X_j$ is at most $K-1$. Hence, the number of clients $N_C$ served by that transmission $X_i + X_j$ is at most $2K-2$. Therefore, an uncoded message would serve a larger number of clients than a coded message when $N_U > N_C$, which means  
    $K \leq \frac{C+2}{3}. $ 
\end{IEEEproof}

\begin{theorem}
\label{thm:mlessc}
    The CDPIC($S$,$K$) coding scheme which can serve $C$ clients when $C = M$, will satisfy the demands of $C$ clients when $C > M$, where $M$ is the number of messages.
\end{theorem}
\begin{IEEEproof}
    Consider the CDPIC($S$,$K$) coding scheme that serves the demands of $C$ clients where $C = M$.  Any client $C_{M+i}$ has the same side information set as that of $C_{i}.$ Since $C_i$ is satisfied by the algorithm for $M=C,$ $C_{M+i}$ will also be satisfied.
\end{IEEEproof}


\begin{table*}[!t]
\caption{Summary of the number of transmissions required for different ranges of cardinality of side information set for CDPIC ($S$, $K$) for $M \leq C.$}
\label{tab:mytable}
\centering
\small
\begin{tabular}{|c|c|c|}
\hline
\textbf{Case}      &\textbf{Transmissions ($N$)}  & \textbf{Whether Optimal?}          \\ \hline
$K \leq \left\lfloor \frac{M}{S+1} \right\rfloor$       & $S+1$     & Yes          \\ \hline
\begin{tabular}[c]{@{}c@{}}$ (n-1) \lfloor \frac{M}{S+n-1} \rfloor < K \leq n \lfloor \frac{M}{S+n} \rfloor\leq \lfloor \frac{M+2}{3} \rfloor $,\\  where $n \in \{2, 3, \ldots, M-S\}$\end{tabular}
 & $ S+n$   & Yes         \\ \hline
$ \lfloor \frac{M+2}{3} \rfloor < K < \lfloor\frac{M}{2}\rfloor$  &$ \begin{cases}2 & S = 1\\\geq S+2 & S > 1\end{cases} $

& 
$ \begin{cases}\text{Optimal}  & S = 1\\ \text{Suboptimal} & S > 1\end{cases} $\\ \hline
$K \geq \left\lfloor \frac{M}{2} \right\rfloor$   & $S+1$  & Yes                                  \\ \hline
\end{tabular}

\end{table*}


\begin{theorem}
\label{Thm:achievability}
    The total number of transmissions $N = S+1$ is achievable for CDPIC($S$,$K$) scheme when $K \leq \left\lfloor \frac{M}{S+1} \right\rfloor$ and $K > \left\lfloor \frac{M}{2} \right\rfloor$, where $M$ is the number of messages.
\end{theorem}
The proof of Theorem \ref{Thm:achievability} and achievable schemes for other ranges of $K$ are elaborated in Section \ref{sec:code}. The important results obtained by code construction in Section \ref{sec:code} is summarized in Table \ref{tab:mytable}. 

The main results in this paper also include the simulation results that demonstrate the increase in convergence rate and accuracy of FedAvg and CELL schemes elaborated in Section \ref{sec:fedavg_appl}. It is observed that the DPIC scheme improves the accuracy by a large margin in general. This is summarized in Table \ref{Tab:Summary_simulation}.

\section{Code construction for CDPIC($S$,$K$)}
\label{sec:code}
This section discusses the code construction for CDPIC ($S$, $K$) problems. Initially, we develop solutions for the case where the number of clients, $C$ is the same as the number of messages, $M$ ($M = C$). By Theorem \ref{thm:mlessc} we know that solutions for these problems would be valid when the number of clients increases above $M$ ($M \leq C$). The case $M \geq C$ is discussed in Section \ref{MgrthC}. 
Since the code construction is different for different ranges of values of $K$, we discuss them separately in the subsequent subsections for the $M=C$ case. For small values of $K$, the optimal code construction is using uncoded transmissions, and for higher values of $K$, coded transmissions are used (Theorem \ref{Thm:uncoded}). 

\subsection{Transmission scheme for the case $K \leq \left\lfloor \frac{M}{S+1} \right\rfloor$}
\label{K_less_M2_a}
The symbols to be transmitted are uncoded in this case. The $S+1$ transmissions are listed below:

\begin{equation}
\label{eq:trans1}
C_{j\lfloor \frac{M}{S+1} \rfloor} \text{ transmits } W_j: X_{1+ j\lfloor \frac{M}{S+1} \rfloor} \text{ for } j \in \{0, 1, \ldots, S\}.
\end{equation}
All the indices are  considered modulo $C.$ 
\begin{lemma}
    Using the transmission in \eqref{eq:trans1}, each client gets $S$ new messages.
\end{lemma}
\begin{IEEEproof}
    The set of $S+1$ transmitted messages is $\{ X_1,X_{1+\lfloor\frac{M}{S+1}\rfloor} ,\ldots ,X_{1+S\lfloor\frac{M}{S+1}\rfloor} \}$. Since $K \leq \left\lfloor \frac{M}{S+1} \right\rfloor,$ no client can possess two or more messages from the set of transmitted messages. So any given client can receive at least $S$ new messages from the transmissions.  
\end{IEEEproof}
Since this meets the lower bound in Theorem \ref{thm:lowerbound}, this transmission scheme is optimal.

\begin{example}
\label{Ex1}
Let's examine a system consisting of 12 messages ($M=12$) and 12 clients ($C=12$), where each client has 3 messages ($K=3$) stored in its side information cache, and each client aims to obtain 3 new messages ($S=3$).
It satisfies the condition $K \leq \lfloor\frac{M}{S+1}\rfloor.$
 
The uncoded symbols transmitted as per \eqref{eq:trans1} are 
    $$W_0=X_1, W_1=X_4, W_2=X_7,
    W_3=X_{10}.$$
The detailed decoding procedure at each client is elaborated in Table \ref{tab1}. In this particular example, each client decodes exactly $S=3$ messages.

\begin{table}[!t]
\centering
\caption{Decoding procedure for Example \ref{Ex1}}
\label{tab1}
\begin{tabular}{| c | c | c | c | c | c | }
\hline
\multicolumn{6}{|c|}{\textbf{Decoding procedure at clients}} \\
\hline
\textbf{client} & \textbf{side information} & \textbf{$W_1$} & \textbf{$W_2$} & \textbf{$W_3$} & \textbf{$W_4$}  \\
\hline
$C_0$ & \{$X_1,X_2,X_{3}\}$& $-$ & $X_4 $ & $X_7$ & $X_{10}$ \\
\hline
$C_1$ &  \{$X_2,X_3, X_{4}\}$ & $X_{1}$ & $-$& $X_7$ & $X_{10}$  \\
\hline
$C_2$ &  \{$X_3,X_4, X_5\}$ & $X_{1}$ & $-$ & $X_7$ & $X_{10}$ \\
\hline
$C_3$ &  \{$X_4,X_5, X_6\}$ & $X_1$ & $-$ & $X_7$ & $X_{10}$ \\
\hline
$C_4$ & \{$X_5, X_6, X_{7}\}$ & $X_1$ & $X_4$ & $-$ & $X_{10}$ \\
\hline
$C_5$ &  \{$X_6, X_7, X_{8}\}$ & $X_1$ & $X_4$ & $-$ & $X_{10}$ \\
\hline
$C_6$ &  \{$X_7, X_8, X_{9}\}$ & $X_1$ & $X_4$& $-$ & $-$  \\
\hline
$C_7$ & \{$X_8, X_9, X_{10}\}$ & $X_1$ & $X_4$ & $X_7$ & $-$ \\
\hline
$C_8$ & \{$X_9,X_{10}, X_{11}\}$ & $X_{1}$ & $X_{4}$& $X_7$ & $-$  \\
\hline
$C_9$ & \{$X_{10},X_{11}, X_0\}$ & $X_1$ & $X_4$ & $X_7$ & $X_{10}$ \\
\hline
$C_{10}$ & \{$X_{11}, X_0, X_1\}$ & $-$ & $X_4$ & $X_7$ & $X_{10}$ \\
\hline
$C_{11}$ & \{$X_{0},X_1, X_2\}$ & $-$ & $X_{4}$ & $X_7$ & $X_{10}$ \\
\hline
\end{tabular}
\end{table}

\end{example}

\subsection{Transmission scheme for the case $(n-1) \lfloor \frac{M}{S+n-1} \rfloor < K \leq n \lfloor \frac{M}{S+n} \rfloor \leq \lfloor \frac{M+2}{3} \rfloor$, where $n \in \{2, 3, \ldots, M-S\}$}.
\label{K_less_M2}

Following from the previous case, we propose an uncoded transmission scheme that requires $S+n$ number of transmissions. 
The transmissions in this case are as follows: 
The client $C_{j\lfloor \frac{M}{S+n} \rfloor} \text{ transmits } $
\begin{equation}
\label{eq:trans2}
 W_j: X_{1+ j\lfloor \frac{M}{S+n} \rfloor} \text{ for } j \in \{0, 1, \ldots, S+n-1\}.
\end{equation}
All the indices are  considered modulo $C.$ 

\begin{lemma}
    Using the transmission in \eqref{eq:trans2}, each client gets $S$ new messages.
\end{lemma}
\begin{IEEEproof}
    The set of $S+n$ transmitted messages is $\{ X_1,X_{1+\lfloor\frac{M}{S+n}\rfloor} ,\ldots, X_{1+(S+n-1)\lfloor\frac{M}{S+n}\rfloor} \}$. The separation between two successive messages in the set of messages transmitted is $\lfloor\frac{M}{S+n}\rfloor$. Since $K \leq n \lfloor \frac{M}{S+n} \rfloor$, no client has $n+1$ or more messages from the uncoded transmissions. 
This guarantees that every client decodes at least $S$ new messages.
\end{IEEEproof}
\begin{lemma}
 The transmission scheme in \eqref{eq:trans2} is optimal. 
\label{lem:S+n}
\end{lemma}
\begin{IEEEproof}
For uncoded transmissions regime, Theorem \ref{thm:basic}
can be rewritten as 
   $$ (M-K)N \geq MS.$$
This follows from the fact that each uncoded message is new for $M-K$ clients. Applying $ K > (n-1)\frac{M}{S+n-1}$, we get
\begin{align*}
    (M-(n-1)\frac{M}{S+n-1})N  &> MS, \\
    \implies N &>S+n-1.
\end{align*}
\end{IEEEproof}

Thus, from Theorem \ref{Thm:uncoded} and Lemma \ref{lem:S+n}, even for this case, the proposed algorithm is optimal with $S+n$ transmissions. This is illustrated in the following example.

\begin{example}
\label{Ex2}
Let's examine a system consisting of 12 messages ($M=12$) and 12 clients ($C=12$), where each client has 4 messages ($K=4$) stored in its side information cache, and each client aims to obtain 4 new messages ($S=4$).
It satisfies the condition $\lfloor\frac{M}{S+1}\rfloor < K \leq 2 \lfloor\frac{M}{S+2}\rfloor.$ Hence $S+2$ transmissions are optimal.
 
The uncoded symbols transmitted as per \eqref{eq:trans2} are 
   $W_0=X_1, W_1=X_3, W_2=X_5,
W_3=X_7,W_4=X_9, W_5=X_{11}$.
The detailed decoding procedure at each client is elaborated in Table \ref{tab2}. In this particular example, each client decodes exactly $S=4$ new messages. Here, the number of broadcast transmissions required to satisfy all the clients with 4 messages is 6.

\begin{table}[!t]
    \centering
    \caption{Decoding procedure for Example \ref{Ex2}}
    \label{tab2}
    \setlength{\tabcolsep}{3.8pt}
     \begin{tabular}{| c | c | c | c | c | c | c | c |}
    \hline
    \multicolumn{8}{|c|}{\textbf{Decoding procedure at clients}} \\
    \hline
    \textbf{Client} & \textbf{Side Information} & \textbf{$W_1$} & \textbf{$W_2$} & \textbf{$W_3$} & \textbf{$W_4$} & \textbf{$W_5$} & \textbf{$W_6$}  \\
    \hline
    $C_0$ & \{$X_1,X_2,X_{3},X_4\}$ & $-$ & $-$ & $X_5$ & $X_7$ & $X_9$ & $X_{11}$\\
    \hline
    $C_1$ & \{$X_2,X_3, X_{4},X_5\}$ & $X_{1}$ & $-$ & $-$ & $X_7$ & $X_9$ & $X_{11}$ \\
    \hline
    $C_2$ & \{$X_3,X_4, X_5,X_6\}$ & $X_{1}$ & $-$ & $-$ & $X_7$ & $X_9$& $X_{11}$ \\
    \hline
    $C_3$ & \{$X_4,X_5, X_6,X_7\}$ & $X_1$ & $X_3$ & $-$ & $-$ & $X_9$& $X_{11}$ \\
    \hline
    $C_4$ & \{$X_5, X_6, X_{7},X_8\}$ & $X_1$ & $X_3$ & $-$ & $-$ & $X_9$ & $X_{11}$\\
    \hline
    $C_5$ & \{$X_6, X_7, X_{8},X_9\}$ & $X_1$ & $X_3$ & $X_5$ & $-$ & $-$ & $X_{11}$ \\
    \hline
    $C_6$ & \{$X_7, X_8, X_{9},X_{10}\}$ & $X_1$ & $X_3$ & $X_5$ & $-$ & $-$ & $X_{11}$ \\
    \hline
    $C_7$ & \{$X_8, X_9, X_{10},X_{11}\}$ & $X_1$ & $X_3$ & $X_5$ & $X_7$ & $-$&$-$ \\
    \hline
    $C_8$ & \{$X_9,X_{10}, X_{11},X_0\}$ & $X_1$ & $X_3$ & $X_{5}$& $X_7$ & $-$  &$-$\\
    \hline
    $C_9$ & \{$X_{10},X_{11}, X_0,X_1\}$ & $-$& $X_3$ & $X_5$ & $X_7$ & $X_9$& $-$\\
    \hline
    $C_{10}$ & \{$X_{11}, X_0, X_1,X_2\}$ & $-$& $X_3$ & $X_5$ & $X_7$ & $X_9$ & $-$\\
    \hline
    $C_{11}$ & \{$X_{0},X_1, X_2,X_3\}$ & $-$& $-$ & $X_{5}$ & $X_7$ & $X_9$ & $X_{11}$\\
    \hline
   \end{tabular}
\end{table}

\end{example}

\subsection{Transmission scheme for the case $ \lfloor \frac{M+2}{3} \rfloor < K < \lfloor\frac{M}{2}\rfloor$:}
\label{K_less_M2_b}
Since $K>\lfloor \frac{M+2}{3} \rfloor, $
single uncoded transmissions are not optimal according to Theorem \ref{Thm:uncoded}. Transmissions in this scenario are as follows: For $S>1,$ $C_{j(M-2K+1)}$ transmits 

\begin{equation}
\label{eq:trans3}
 W_j : X_{1+j(M-2K+1)} \oplus X_{K+j(M-2K+1)}, \text{for}  j \in \{0, 1, \ldots\}.
\end{equation}

For $S=1$, only two transmissions suffice, i.e.,
$$C_0 \text{ transmits } X_1 \oplus X_{K} \text{ and }$$ 
$$ C_{M-2K+1} \text{ transmits } X_{M-2K+2} \oplus X_{M-K+1}.  $$

\begin{lemma}
\label{lem:4}
Using the transmission in \eqref{eq:trans3}, each client gets $S$ new messages from the transmitted coded symbols.
\end{lemma}

\begin{IEEEproof}
For $S=1,$ by the first transmission $X_1+X_K,$ a total of $2(K-1)$ clients get satisfied with their demands. The $K-1$ clients immediate next to $C_0$, i.e., $C_1, C_2, \ldots, C_{K-1}$ have $X_K$ in their memory and hence can decode $X_1.$ Similarly the last $K-1$ clients $C_{M-K-2}, C_{M-K-1}, \ldots, C_{M-1}$ decode $X_K$ from the first transmission. The clients that are left out without decoding are transmitter $C_0$ and $C_{K}, C_{K+1}, \ldots, C_{M-K}.$ 
From the second transmission, $X_{M-2K+2} \oplus X_{M-K+1},$ these clients also get what they request.

For $S>1,$ we can continue with the transmission scheme used in $S=1.$ But, since there can be repetitions in the decoded messages and some receivers not being able to decode from more than one transmission, we would require more than $S+1$ number of transmissions.

\end{IEEEproof}

Here, the transmission algorithm mentioned above is optimal for \( S = 1 \), but it is in general sub-optimal for \( S \geq 2 \).

This is illustrated in the following examples.
\begin{example}
\label{Ex3}
Consider a system comprising 16 messages (\(M=16\)) and 16 clients (\(C=16\)). In this setup, each client possesses 7 messages (\(K=7\)) stored in its side information cache, while aiming to acquire 2 new messages (\(S=1\)).

Here $\lfloor \frac{M+2}{3} \rfloor < K < \lfloor\frac{M}{2}\rfloor $, so from equation \eqref{eq:trans3} and Lemma \ref{lem:4} the coded symbols of transmitting clients set $\{C_0,C_3\}$ are $X_{1}\oplus X_{7},X_{4}\oplus X_{10}$ respectively.

\begin{table}[!t]
    \centering
    \caption{Decoding procedure for Example \ref{Ex3}}
    \label{tab3} 
     \begin{tabular}{| c | c | c |}
\hline
\multicolumn{3}{|c|}{\textbf{Decoding procedure at clients}} \\
\hline
\textbf{client} & \textbf{$W_1$} & \textbf{$W_2$}   \\
\hline
$C_0$  \{$X_1, \ldots, X_7$\} & $-$ & $X_{10}$  \\
\hline
$C_1$   \{$X_2, \ldots, X_8$\} & $X_{1}$ & $X_{10}$ \\
\hline
$C_2$   \{$X_3, \ldots, X_9$\} & $X_{1}$ & $X_{10}$ \\
\hline
$C_3$   \{$X_4, \ldots , X_{10}$\} & $X_{1}$ & $-$  \\
\hline
$C_4$  \{$X_5, \ldots, X_{11}$\} & $X_{1}$ & $X_4$ \\
\hline
$C_5$   \{$X_6, \ldots, X_{12}$\} & $X_{1}$ & $X_4$  \\
\hline
$C_6$   \{$X_7, \ldots, X_{13}$\} & $X_1$ & $X_4$    \\
\hline
$C_7$  \{$X_8, \ldots, X_{14}$\} & $-$ & $X_4$   \\
\hline
$C_8$  \{$X_9, \ldots, X_{15}$\} & $-$ & $X_4$ \\
\hline
$C_9$  \{$X_{10}, \ldots, X_{0}$\} & $-$ & $X_4$   \\
\hline
$C_{10}$ \{$X_{11}, \ldots, X_{1}$\} & $X_7$ & $-$   \\
\hline
$C_{11}$ \{$X_{12}, \ldots, X_{2}$\} & $X_7$ & $-$  \\
\hline
$C_{12}$ \{$X_{13}, \ldots, X_{3}$\} & $X_7$ & $-$  \\
\hline
$C_{13}$ \{$X_{14}, \ldots, X_{4}$\} & $X_7$ & $X_{10}$ \\
\hline
$C_{14}$ \{$X_{15}, \ldots, X_{5}$\} & $X_7$ & $X_{10}$   \\
\hline
$C_{15}$ \{$X_{0}, \ldots, X_{6}$\} & $X_7$ & $X_{10}$   \\
\hline
\end{tabular}
 \end{table}
The detailed decoding procedure for all clients is elaborated in Table \ref{tab3}. In this particular example, each client decodes \( S=1 \) new message using two transmissions.
\end{example}

\begin{example}
\label{Ex4}
Consider a system comprising 16 messages (\(M=16\)) and 16 clients (\(C=16\)). In this setup, each client possesses 7 messages (\(K=7\)) stored in its side information cache, while aiming to acquire 2 new messages (\(S=2\)).

Here $\lfloor \frac{M+2}{3} \rfloor < K < \lfloor\frac{M}{2}\rfloor $, so from equation \eqref{eq:trans3} and Lemma \ref{lem:4} the coded symbols of transmitting clients set $\{C_0,C_3,C_{6},C_9\}$ are $X_{1}\oplus X_{7},X_{4}\oplus X_{10},X_{7}\oplus X_{13},X_{10}\oplus X_{0}$ respectively.
\begin{table}[!t]
    \centering
    \caption{Decoding procedure for Example \ref{Ex4}}
     \begin{tabular}{| c | c | c | c | c |}
\hline
\multicolumn{5}{|c|}{\textbf{Decoding procedure at clients}} \\
\hline
\textbf{client} & \textbf{$W_1$} & \textbf{$W_2$} & \textbf{$W_3$} & \textbf{$W_4$}  \\
\hline
$C_0$  \{$X_1, \ldots, X_7$\} & $-$ & $X_{10}$ & $X_{13}$ & $-$\\
\hline
$C_1$   \{$X_2, \ldots, X_8$\} & $X_{1}$ & $X_{10}$ & $X_{13}$ & $-$ \\
\hline
$C_2$   \{$X_3, \ldots, X_9$\} & $X_{1}$ & $X_{10}$  & $X_{13}$ & $-$ \\
\hline
$C_3$   \{$X_4, \ldots , X_{10}$\} & $X_{1}$ & $-$  & $X_{13}$ & $X_0$ \\
\hline
$C_4$  \{$X_5, \ldots, X_{11}$\} & $X_{1}$ & $X_4$  & $X_{13}$  & $X_0$ \\
\hline
$C_5$   \{$X_6, \ldots, X_{12}$\} & $X_{1}$ & $X_4$  & $X_{13}$  & $X_0$ \\
\hline
$C_6$   \{$X_7, \ldots, X_{13}$\} & $X_1$ & $X_4$  & $-$   & $X_0$\\
\hline
$C_7$  \{$X_8, \ldots, X_{14}$\} & $-$ & $X_4$  & $X_7$   & $X_0$\\
\hline
$C_8$  \{$X_9, \ldots, X_{15}$\} & $-$ & $X_4$  & $X_7$  & $X_0$ \\
\hline
$C_9$  \{$X_{10}, \ldots, X_{0}$\} & $-$ & $X_4$   & $X_7$  & $-$ \\
\hline
$C_{10}$ \{$X_{11}, \ldots, X_{1}$\} & $X_7$ & $-$  & $X_7$ & $X_{10}$ \\
\hline
$C_{11}$ \{$X_{12}, \ldots, X_{2}$\} & $X_7$ & $-$  & $X_7$ & $X_{10}$ \\
\hline
$C_{12}$ \{$X_{13}, \ldots, X_{3}$\} & $X_7$ & $-$  & $X_7$ & $X_{10}$  \\
\hline
$C_{13}$ \{$X_{14}, \ldots, X_{4}$\} & $X_7$ & $X_{10}$  & $-$  & $X_{10}$ \\
\hline
$C_{14}$ \{$X_{15}, \ldots, X_{5}$\} & $X_7$ & $X_{10}$  & $-$ & $X_{10}$ \\
\hline
$C_{15}$ \{$X_{0}, \ldots, X_{6}$\} & $X_7$ & $X_{10}$  & $-$  & $X_{10}$\\
\hline
\end{tabular}
 \label{tab4} 
\end{table}

The detailed decoding procedure for all clients is elaborated in Table \ref{tab4}. In this particular example, each client decodes  $S=2$ new messages from four transmissions.
\end{example}

\subsection{Transmission scheme for the case $K\in \{\lfloor\frac{M}{2}\rfloor,\lfloor\frac{M}{2}\rfloor+1$\}:}
\label{K_eqM2}
In this scenario, where each client holds either $\lfloor\frac{M}{2}\rfloor \text{ or }\lfloor\frac{M}{2}\rfloor+1$ messages as side information, the coded symbols in the proposed transmission consist of two messages: Client $C_j$ transmits 
\begin{equation}
\label{eq:trans4}
X_{j+1} \oplus X_{j+K}, \text{ where } j \in \{0, K-2, K-3, \ldots, K-S-1\}.    
\end{equation}

\begin{lemma}
\label{lem:5}
    Using the transmission in \eqref{eq:trans4}, each client gets $S$ new messages.
\end{lemma}
\begin{IEEEproof}
We see that each transmission in \eqref{eq:trans4} corresponds to the transmission of the first and last message in the side information set. Since the number of side information messages is $\lfloor\frac{M}{2}\rfloor \text{ or }\lfloor\frac{M}{2}\rfloor+1$, from the first transmission, except for the client that transmits ($C_0$), all other clients get a new message. Since the next transmission is made by client $C_{K-2},$ it is ensured that the transmitted message $X_{K-1} \oplus X_{2K-2}$ satisfies $C_0$ too, as one message in the coded symbol is in the side-information set of $C_0.$ Hence the algorithm holds for $S=1$ with $S+1=2$ transmissions. 

Assume that the algorithm holds for $S=s,$ which means that the algorithm satisfies all the clients with $s+1$ transmissions. The transmissions are made by the clients $C_0, C_{K-2}, C_{K-3}, \ldots, C_{K-s-1}.$ The next transmission is made by $C_{K-s}$ and the transmission is $X_{K-s+1} \oplus X_{2K-s}.$
Client $C_{K-s}$ has not transmitted before. Hence, it has already received $s+1$ new messages from the previous transmissions. The transmitted coded message includes messages that have not been transmitted so far. Hence, other clients get a new message from this particular transmission. Hence, the decodability is proved by using induction.
\end{IEEEproof}
Since this meets the lower bound in Theorem \ref{thm:lowerbound}, this transmission scheme is optimal. Example \ref{Ex5} illustrates this algorithm.
 \begin{example}
\label{Ex5}
Consider a system comprising 11 messages ($M=11$) and 11 clients ($C=11$). Each client possesses a cache of side information containing 6 messages ($K=6$) and aims to acquire 5 new messages ($S=5$).

    From equation \eqref{eq:trans4} and Lemma \ref{lem:5} the set of clients which are satisfy all the clients with $S+1$ transmissions are $\{C_5,C_9,C_8,C_6,C_6,C_4\}$ and respective coded symbols are $ X_6 \oplus X_{0}, X_{10} \oplus X_4, X_9 \oplus X_3, X_8 \oplus X_2, X_7 \oplus X_1, X_7 \oplus X_1, X_5 \oplus X_{10}.$

The detailed decoding procedure for all clients is elaborated in Table \ref{tab5}. In this particular example, each client decodes exactly $S=3$ messages.

\begin{table}[!t]
\centering
\caption{Decoding procedure for Example \ref{Ex5}}
\label{tab5}
\begin{tabular}{| c | c | c | c | c | c | c |}
\hline
\multicolumn{7}{|c|}{\textbf{Decoding procedure at clients}} \\
\hline
\textbf{client} & \textbf{$W_1$} & \textbf{$W_2$} & \textbf{$W_3$} & \textbf{$W_4$} & \textbf{$W_5$} & \textbf{$W_6$}  \\
\hline
$C_0$  \{$X_1, \ldots, X_6$\} & $X_{0}$ & $X_{10}$ & $X_9$ & $X_8$ & $X_7$  & $X_{10}$\\
\hline
$C_1$   \{$X_2, \ldots, X_7$\} & $X_{0}$ & $X_{10}$ & $X_9$ & $X_8$ & $X_1$ & $X_{10}$\\
\hline
$C_2$   \{$X_3, \ldots, X_8$\} & $X_{0}$ & $X_{10}$ & $X_9$ & $X_2$ & $X_1$ & $X_{10}$ \\
\hline
$C_3$   \{$X_4, \ldots , X_9$\} & $X_{0}$ & $X_{10}$ & $X_3$ & $X_2$ & $X_1$  & $X_{10}$ \\
\hline
$C_4$  \{$X_5, \ldots, X_{10}$\} & $X_{0}$ & $X_4$  & $X_3$ & $X_2$ & $X_1$ & $-$ \\
\hline
$C_5$   \{$X_6, \ldots, X_{0}$\} & $-$ & $X_4$ & $X_3$ & $X_2$ & $X_1$  & $X_5$ \\
\hline
$C_6$   \{$X_7, \ldots, X_{1}$\} & $X_6$ & $X_4$  & $X_3$ & $X_2$ & $-$ & $X_5$ \\
\hline
$C_7$  \{$X_8, \ldots, X_{2}$\} & $X_6$ & $X_4$ & $X_3$ & $-$ & $X_7$  & $X_5$ \\
\hline
$C_8$  \{$X_9, \ldots, X_{3}$\} & $X_{6}$ & $X_4$ & $-$ & $X_8$ & $X_7$  & $X_5$ \\
\hline
$C_9$  \{$X_{10}, \ldots, X_{4}$\} & $X_6$ & $-$  & $X_9$ & $X_8$ & $X_7$ & $X_5$ \\
\hline
$C_{10}$  \{$X_{0}, \ldots, X_{5}$\} & $X_6$ & $X_{10}$  & $X_9$ & $X_8$ & $X_7$ & $-$ \\
\hline

\end{tabular}
\end{table}
\end{example}

In Example \ref{Ex5} provided earlier, all clients received $S$ new messages through $S+1$ transmissions.

\subsection{Cardinality of each client $K > \frac{M}{2}+1$:}
\label{K_grM2}
Since the cardinality of the side-information set is on the higher side, the transmission scheme involves coded transmissions that encode more than two messages. The transmission algorithm is elaborated below.

Each transmission involves encoding of $e+1$ messages, where $e$ is defined as $e=K-\lfloor\frac{M}{2} \rfloor.$ The messages that are encoded should be maximally separated. Hence, the scheme is: $C_j$ transmits 
\begin{equation}
\label{eq:trans5}
\bigoplus_{i=0}^{e}  X_{j+1+i\lfloor \frac{K}{e} \rfloor}   
\end{equation}
for $j\in \{0,1, \ldots, S\}.$

\begin{lemma}  Using the transmission in \eqref{eq:trans5}, each client gets $S$ new messages with optimum number of transmissions i.e., $N=S+1$.
\end{lemma}

   \begin{IEEEproof}
Since $e>1,$ any client that possess $X_i$ has either $X_{i+\lfloor\frac{K}{e} \rfloor}$ or $X_{i-\lfloor\frac{K}{e} \rfloor}$ as side information. Since this is true for any $i$, any given client has all except one message in this coded symbol in its side information set. Since consecutive clients are transmitting, at least the first and last messages in the coded symbol are distinct for each transmission. Also, for the transmitting clients, there is no repeated decoding of the same message since the first message in any transmission is unknown to all other transmitting clients. 
\end{IEEEproof}

\begin{example}
\label{Ex6}
Let's examine a system consisting of 10 messages ($M=10$) and 10 clients ($C=10$), where each client has 7 messages ($K=7$) stored in its side information cache, and each client aims to obtain 3 new messages ($S=3$).

 According to the algorithm mentioned above, the coded symbols for the given system, to satisfy all clients with $S+1$ transmissions, would be:
\begin{center}

From client $C_0$: $W_1 = X_1 \oplus X_4 \oplus X_7,$ \\
From client $C_1$: $W_2 = X_2 \oplus X_5 \oplus X_8,$ \\
From client $C_2$: $W_3 = X_3 \oplus X_6 \oplus X_9,$ \\
From client $C_3$: $W_4 = X_4 \oplus X_7 \oplus X_0.$ \\

\end{center}

The detailed decoding procedure at each client is elaborated in Table \ref{tab6}. In this particular example, each client decodes exactly $S=3$ messages.

\begin{table}[!t]
    \centering
    \caption{Decoding procedure for Example \ref{Ex6}}
    \begin{tabular}{| c | c | c |c | c |}
\hline
\multicolumn{5}{|c|}{\textbf{Decoding procedure at clients}} \\
\hline
\textbf{client} & \textbf{$W_1$} & \textbf{$W_2$} & \textbf{$W_3$} & \textbf{$W_4$} \\
\hline
$C_0$  \{$X_1,X_2 \ldots, X_7$\} & $-$ & $X_{8}$ & $X_9$ & $X_0$ \\
\hline
$C_1$   \{$X_2, X_3 \ldots, X_8$\} & $X_{1}$ & $-$ & $X_9$ & $X_0$ \\
\hline
$C_2$   \{$X_3, X_4 \ldots, X_9$\} & $X_{1}$ & $X_{2}$ & $-$ & $X_0$ \\
\hline
$C_3$   \{$X_4,X_5 \ldots , X_0$\} & $X_{1}$ & $X_{2}$ & $X_3$ & $-$ \\
\hline
$C_4$  \{$X_5, X_6\ldots, X_{1}$\} & $X_{4}$ & $X_2$  & $X_3$ & $X_4$ \\
\hline
$C_5$   \{$X_6, X_7 \ldots, X_{2}$\} & $X_4$ & $X_5$ & $X_3$ & $X_4$ \\
\hline
$C_6$   \{$X_7, X_8\ldots, X_{3}$\} & $X_4$ & $X_5$  & $X_6$ & $X_4$ \\
\hline
$C_7$  \{$X_8, X_9 \ldots, X_{4}$\} & $X_7$ & $X_5$ & $X_6$ & $X_7$ \\
\hline
$C_8$  \{$X_9, X_0\ldots, X_{5}$\} & $X_{7}$ & $X_8$ & $X_6$ & $X_7$ \\
\hline
$C_9$  \{$X_{0}, X_1 \ldots, X_{6}$\} & $X_7$ & $X_8$  & $X_9$ & $X_7$ \\
\hline
\end{tabular}
\label{tab6}
\end{table}
\end{example}
\subsection{CDPIC$(S,K)$ for $M > C$}
\label{MgrthC}
For the case $M>C,$ the following observations could be made. For $K\leq \frac{C+2}{3},$ according to Theorem \ref{Thm:uncoded}, uncoded transmissions are optimal. We can use the same uncoded transmissions that we used for the case of $M=C,$ with only the difference that the client that transmits the message would be different from the earlier case.

In the case of coded transmissions, the ones for $K\geq \frac{M}{2},$ the nature of the algorithm in $M=C$ ensures that the same can be used without any change for $M>C$ as well. But for the case $ \lfloor \frac{M+2}{3} \rfloor < K < \lfloor\frac{M}{2}\rfloor,$ the code design would differ and a larger number of transmissions would be required. 

\section{Application of CDPIC($S$,$K$) to  FL in ITS}
\label{sec:fedavg_appl}
This section discusses the experimental setup for 
FL considering non-IID data distribution in ITS. The results indicating improvement in FL performance by applying proposed CDPIC($S$,$K$) solutions are presented with detailed analysis.

\subsection{Experimental setup}
Each client node independently trains a CVAE using its local dataset. The training process involves optimizing a variational lower bound objective that combines reconstruction loss and KL divergence. The encoder and decoder networks are both conditioned on one-hot encoded class labels to ensure structured data generation. Each client trains the CVAE for a maximum of 40 epochs and generates synthetic images that align with its local data distribution. These synthetic data are later utilized for shuffling in the FL setup.

The FedAvg and CELL
 are implemented by using
CNN-based FL with MNIST in pytorch. We tuned the local epochs
as 2 and adjusted the communication round limit to [100, 150, 300]. The CNN model trained with
a learning rate of 0.001 and a batch size of 128. Momentum
0.9 was utilized with the SGD optimizer. Then, after each
communication round, the global model test accuracy is checked with the test dataset. This is repeated for each pliable data transmission case. This simulation is repeated by using Light Weight CNN(LW-CNN)-instead of CNN in the FL setup.

\textbf{Model and Dataset:}  MNIST  contains
the gray-scale images of $6 \times 10^4 $ handwritten digits with size $28 \times 28$  ( $5 \times 10^4$
for training and $10^4$ for testing ) \cite{MNIST}. Similarly, the CIFAR-10 \cite{cifar-10} comprises  colored images of size $32 \times 32$ pixels with RGB channels ( $5 \times 10^4$
for training and $10^4$ for testing ). 

CNN and LW-CNN are the models used inside the FL with MNIST. This CNN model has 421,642 trainable parameters and consists of two convolutional layers with ReLU activation and max pooling, followed by two fully connected layers with dropout for regularization. The efficiency-focused LW-CNN has 4,254 parameters and consists of two convolutional layers (weights only, no bias) and one fully connected layer. The LW-CNN architecture will be suitable for ML applications when the data samples in different classes exhibit similarity among them. For applications such as autonomous driving,  the VANET image samples will have enough similarity, making LW-CNN a suitable architecture. LW-CNN helps to reduce FL transmission overhead, making it appropriate for resource-constrained vehicular edge devices. But with the more complex data set, CNN’s robust architecture will be essential for good results.
The experimental findings obtained using the MNIST are further validated on the CIFAR-10 dataset. To accommodate the differences in input characteristics, the CNN architecture is modified by adapting the input layer for RGB images.
\begin{figure}[!t]
\centering
\includegraphics[width=3.2in]{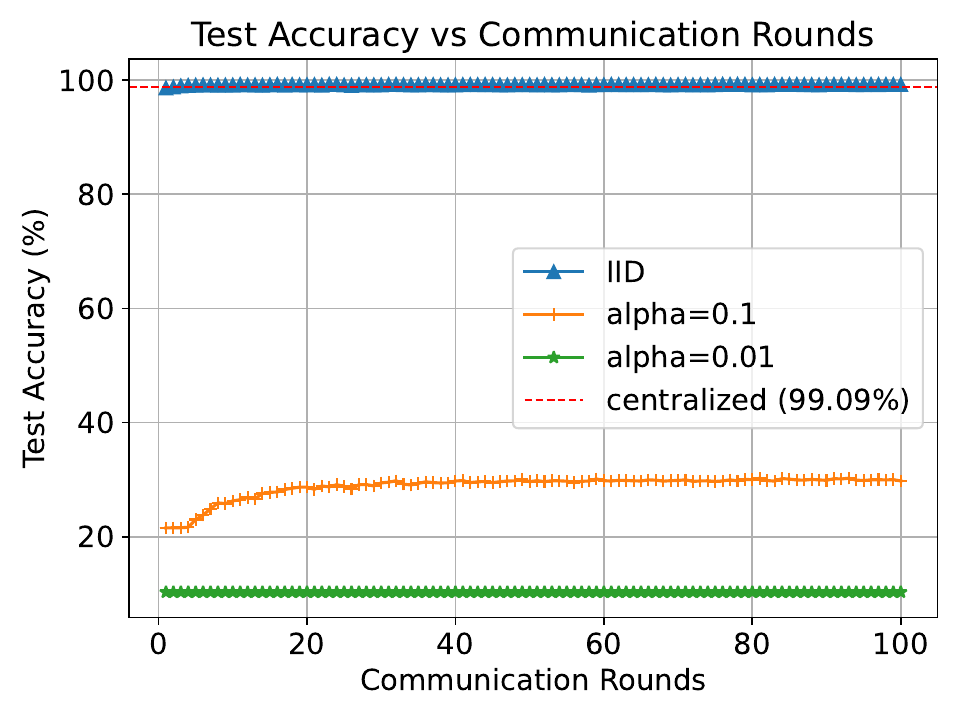}
\caption{Comparison of the performance of CML, FedAvg with IID local data across 5 users, and FedAvg with non-IID data across 5 users in a CNN-based  FL setup with MNIST dataset. Here, alpha($\alpha$) value indicates the degree of non-IID nature in the Dirichlet distribution}
\label{fig_comp}
\end{figure}
\subsection{ Results and Analysis}
\label{sec:results}

In our initial experiments, we trained a CNN on the MNIST dataset using the centralized machine learning (CML) set-up, where it is assumed that the entire database and ML model are available on the central server. Subsequently, as a baseline, we evaluated the performance of FedAvg algorithm using both IID and non-IID data from MNIST. In this study, we generated non-IID data distributions using Dirichlet sampling, guided by a concentration parameter $\alpha$ \cite{li2024synthetic}. We selected 
$\alpha \in \{0.01, 0.1\}$
as these values are commonly employed in similar research \cite{lin2020ensemble}. As $\alpha$ decreases, it increases the likelihood that individual clients will take samples exclusively from a single randomly chosen class.
Fig. \ref{fig_comp} presents a comparative evaluation of CML, FedAvg with IID data, and FedAvg with non-IID data, and the results are summarized in Table \ref{tab6}. 
\begin{table}[!t]
\caption{FedAvg performance with IID/nonIID datasets in MNIST\label{tab6}}

\centering
\begin{tabular}{|c|c|c|}
\hline
\textbf{Datasets} & \textbf{Fedavg Model} & \textbf{Test accuracy} \\ 
\hline
\multirow{3}{*}{MNIST} & Centralized Machine Learning (CML) & 99.09 \\ 
\cline{2-3} 
 & Fedavg with IID & 99.09\\ 
\cline{2-3} 
 & Fedavg with nonIID ($\alpha$=0.1) & 30.00\\
\cline{2-3}
 & Fedavg with nonIID ($\alpha$=0.01) & 15.00\\
\hline
\end{tabular}
\end{table}
Results in Table \ref{tab6} show that FedAvg with IID data at the client nodes demonstrates performance comparable to CML. 
These results also show that the performance of FL degrades significantly when the data is non-IID. This motivates data shuffling among nodes for performance improvement.
\begin{figure}[!t]
    \centering
    \includegraphics[width=3.2in]{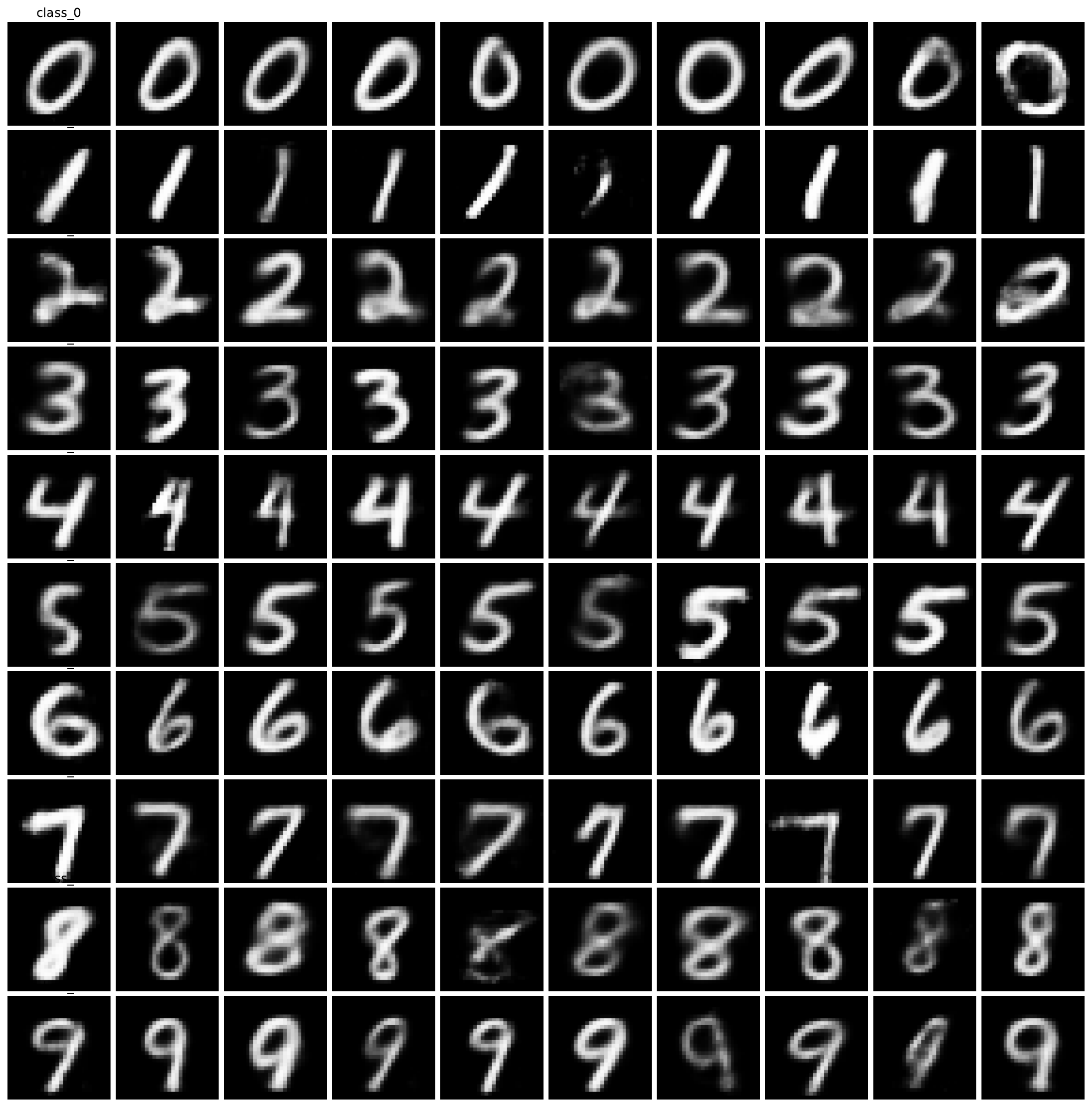}
    \caption{Synthetic data of each class of MNIST by using CVAE}
    \label{fig_syntheticdata}
\end{figure}

\begin{figure}[!t]
    \centering
    \includegraphics[width=3.2in]{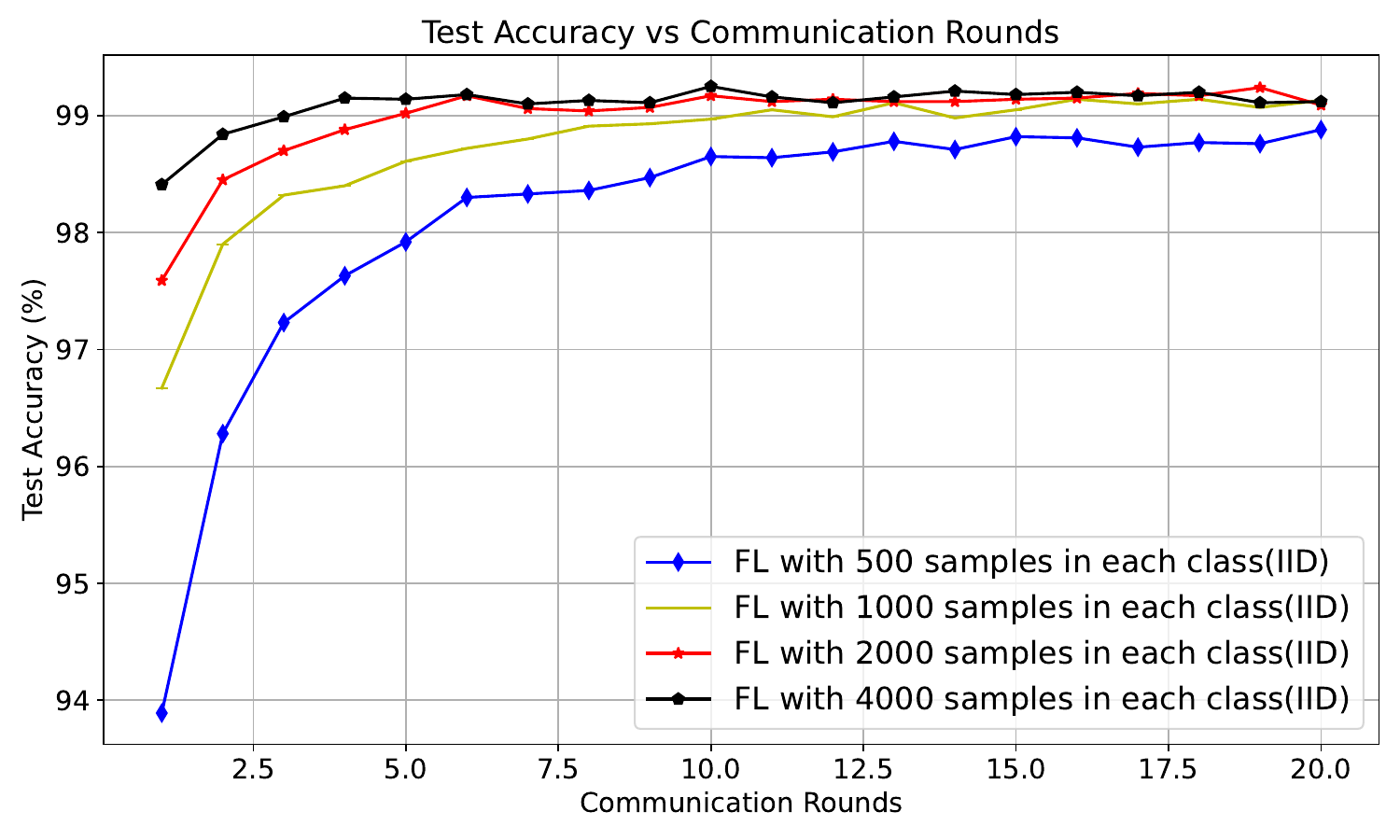}
    \caption{Performance Comparison of CNN-Based FL Using FedAvg in a 10-User IID Setup by Varying the Number of Samples per Class}
    \label{fig_varynumbersamples}
\end{figure}

In our system model, we consider $C$ RSUs as nodes and $M=10$ messages, representing 10 distinct data classes from the MNIST/ CIFAR10 dataset. Each RSU collects local data from vehicles within a specific geographical area, and adjacent RSUs access common geographical regions, giving rise to overlap in their local data samples, making the setting the same as CDPIC($S$,$K$).  Here $K$ is the number of data classes possessed by each node and there is an overlap of $K{-}1$ classes in the side information sets of adjacent nodes. This $K$ is usually less than $M$, making the data non-IID. The non-IID distribution of the data depends upon the parameters $M$,$C$,$K$. As $K$ increases, the overlap of data classes among consecutive RSUs also increases, leading to a more homogeneous data distribution across nodes. 

In our experiment, we considered two scenarios with $K=6$ and $K=7$, corresponding to the cases where $K\in \{\lfloor\frac{M}{2}\rfloor,\lfloor\frac{M}{2}\rfloor+1$\} and $K>\frac{M}{2}+1$, respectively. These settings help us evaluate how pliable data shuffling improves FL performance under varying levels of non-IID data distribution. Due to the simple nature of the MNIST dataset, especially its high similarity between classes, the FL model can achieve close to 90 $\%$ accuracy even without data shuffling when $K > \frac{M}{2}$
. However, for more practical and complex datasets, FL performance degrades significantly under non-IID conditions, as can be clearly seen from Table \ref{Tab:Summary_simulation_CIFAR10}. It can be seen that with the CIFAR 10 dataset, CNN could give only 45 $\%$ accuracy for $M=C=10$ and $K=6$. This degradation arises due to the higher heterogeneity in available data classes and the lack of sufficient class diversity needed to approximate an IID distribution.

When data distribution is non-IID, for improving FL performance, this work proposes to shuffle the synthetically generated samples from the CVAE of each node to other nodes.  Since the CVAE of each node is trained with its local data, it will generate synthetic images corresponding to the classes present in its local dataset. However, since the CVAE reconstructs images from the latent representations of the original data, the generated images are not exact replicas of the local data. Instead, they provide a representative reconstruction that captures the underlying characteristics of the original distribution. Therefore, there could be a slight reduction in the accuracy of FL when trained on synthetic data. Samples of synthetic images generated using CVAE are given in Fig. \ref{fig_syntheticdata}. To simulate the real VANET scenario where the available number of data samples in a class may differ among nodes, our experimental setup assigns each node with different number of samples/class. Further, we apply data augmentation techniques to equalize the number of samples/class within each node.

Different levels of data shuffling achieved through the CDPIC schemes derived in Section \ref{sec:code} help to progressively update the data distribution towards IID. In this work, we specifically evaluate the performance of FL under different levels of data shuffling for non-IID scenarios where $K \geq \frac{M}{2}$. With data shuffling, each node with $K$ data classes may benefit from receiving samples from the remaining $M{-}K$ data classes. How many data samples are to be shuffled for achieving performance close to IID  depends on the complexity of the model, the FL algorithm, and the nature of the dataset. Experiments are conducted to evaluate the impact of proposed CDPIC($S$,$K$) schemes on improving FL accuracy, convergence speed, and transmission overhead. Results are presented through convergence plots and tables.

To achieve performance comparable to centralized training within a few rounds, where all users have IID data, each user in the 10-node FL setup requires at least 2000 samples per class, as illustrated in Fig. \ref{fig_varynumbersamples}.  Under such IID conditions, FL converges to the optimal accuracy within just 5 communication rounds. Therefore, we evaluated FL performance in the fifth round to ensure a fair comparison.
\begin{table*}[h]
\centering
\caption{Transmission Efficiency in FL using CNN with the Proposed CDPIC($S$,$K$) Schemes }
\label{Tab:Summary_simulation}
\begin{adjustbox}{width=0.9\linewidth}
\renewcommand{\arraystretch}{1.3}
\begin{tabular}{|c|c|c|c|c|c|c|}
\hline
Case          & \multicolumn{1}{c|}{\begin{tabular}[c]{@{}c@{}}Number of  \\ required \\ new data\\  classes\end{tabular}} & \multicolumn{1}{c|}{\begin{tabular}[c]{@{}c@{}}Number of \\ transmissions \\ without using \\ CDPIC Scheme ($N_W$)\end{tabular}} & \multicolumn{1}{c|}{\begin{tabular}[c]{@{}c@{}}Number of \\ transmissions \\ with using CDPIC\\  Scheme ($N$)\end{tabular}} & \multicolumn{1}{c|}{\begin{tabular}[c]{@{}c@{}}Transmission\\  Efficiency\\ ($\%$)     \\  $\frac{(N_W-N)}{N_W} \times 100$\end{tabular}} & \multicolumn{2}{c|}{\begin{tabular}[c]{@{}c@{}}Accuracy improvment \\ after 5 rounds \\($\%$)\end{tabular}} \\  \cline{6-7}
\multicolumn{1}{|c|}{} &     &  & & & FedAvg+CNN & CELL+CNN     \\ \hline

\multirow{5}{*}{\begin{tabular}[c]{@{}c@{}}$M=C=10,K=6$  \\Sec. \ref{K_eqM2}        \\ Fig. \ref{Fig_M10C10K6_3}\end{tabular}} & $S=0$      & -  &   - &       -    &  91.00 & 87.00\\ \cline{2-7} 
   & $S=1$ & 3         &   2        &   33.33   & 95.00 &  91.00   \\ \cline{2-7} 
           & $S=2$ &    5      &     3      &     40 &   96.50 & 94.00  \\ \cline{2-7} 
           & $S=3$ &   8       &    4       &    50  & 98.90   & 98.50 \\ \cline{2-7} 
           & $S=4$ &  10        &  5         &   50   &  98.90    &98.50 \\ \hline
\multirow{5}{*}{\begin{tabular}[c]{@{}c@{}}$M=C=10,K=7$\\Sec. \ref{K_grM2} \\       Fig. \ref{Fig_M10C10K7_3}\end{tabular}}     & $S=0$ &   -       & -       -  &   -   &   95.50 & 92.00  \\ \cline{2-7} 
           & $S=1$ &  4        &       2    &    50  &   98.90 &98.50 \\ \cline{2-7} 
           & $S=2$ &     7     &    3       &     57.14 &   98.90 & 98.50  \\ \cline{2-7} 
           & $S=3$ &    10      &     4      &  60    &   98.90 &  98.50  \\ \hline

\end{tabular}
\end{adjustbox}
\end{table*}

\begin{table*}[h]
\centering
\caption{Reduction in Number of Communication Rounds in FL using CNN with Proposed CDPIC($S$,$K$) Schemes}
\label{Tab:Summary}
\begin{adjustbox}{width=0.8\linewidth}
\renewcommand{\arraystretch}{1.3}
\begin{tabular}{|l|c|c|c|c|}
\hline

\multicolumn{1}{|c|}{\multirow{2}{*}{Case}} & \multicolumn{1}{c|}{\begin{tabular}[c]{@{}c@{}}Number of  \\ required new data \\ classes to get \\CML accuracy\end{tabular}}& \multicolumn{3}{c|}{\begin{tabular}[c]{@{}l@{}}\\Number of FL rounds  to reach CML Accuracy \\\end{tabular}}            \\ \cline{3-5}
\multicolumn{1}{|c|}{} &       &  & FedAvg+CNN & CELL+CNN     \\ \hline

\multirow{2}{*}{\begin{tabular}[c]{@{}c@{}}$M = C = 10, K = 6  $           \\  Fig. \ref{Fig_M10C10K6_3}\end{tabular}}        & \multirow{2}{*}{\begin{tabular}[c]{@{}c@{}}$S=3$\end{tabular}} &  \multicolumn{1}{c|}{Initial non-IID Data:}     & \multicolumn{1}{c|}{80}       &  110 \\ \cline{3-5}
 &                            & \multicolumn{1}{c|}{After $S+1$   Transmissions:} & \multicolumn{1}{c|}{10}       &  13 \\ \cline{3-5}

 \hline
\multirow{2}{*}{\begin{tabular}[c]{@{}c@{}}$M = C = 10, K = 7$             \\Fig. \ref{Fig_M10C10K7_3}\end{tabular}} & \multirow{2}{*}{\begin{tabular}[c]{@{}c@{}}$S=1$\end{tabular}}  &  \multicolumn{1}{c|}{Initial non-IID Data:}     & \multicolumn{1}{c|}{63}       &80  \\ \cline{3-5}
&                            & \multicolumn{1}{c|}{After $S+1$ Transmissions:} & \multicolumn{1}{c|}{8}       & 9    
\\ \cline{3-5}

\hline
\end{tabular}

\end{adjustbox}
\end{table*}

\begin{figure*}[!t]
\centering
\subfloat[]{\includegraphics[width=3.2in]{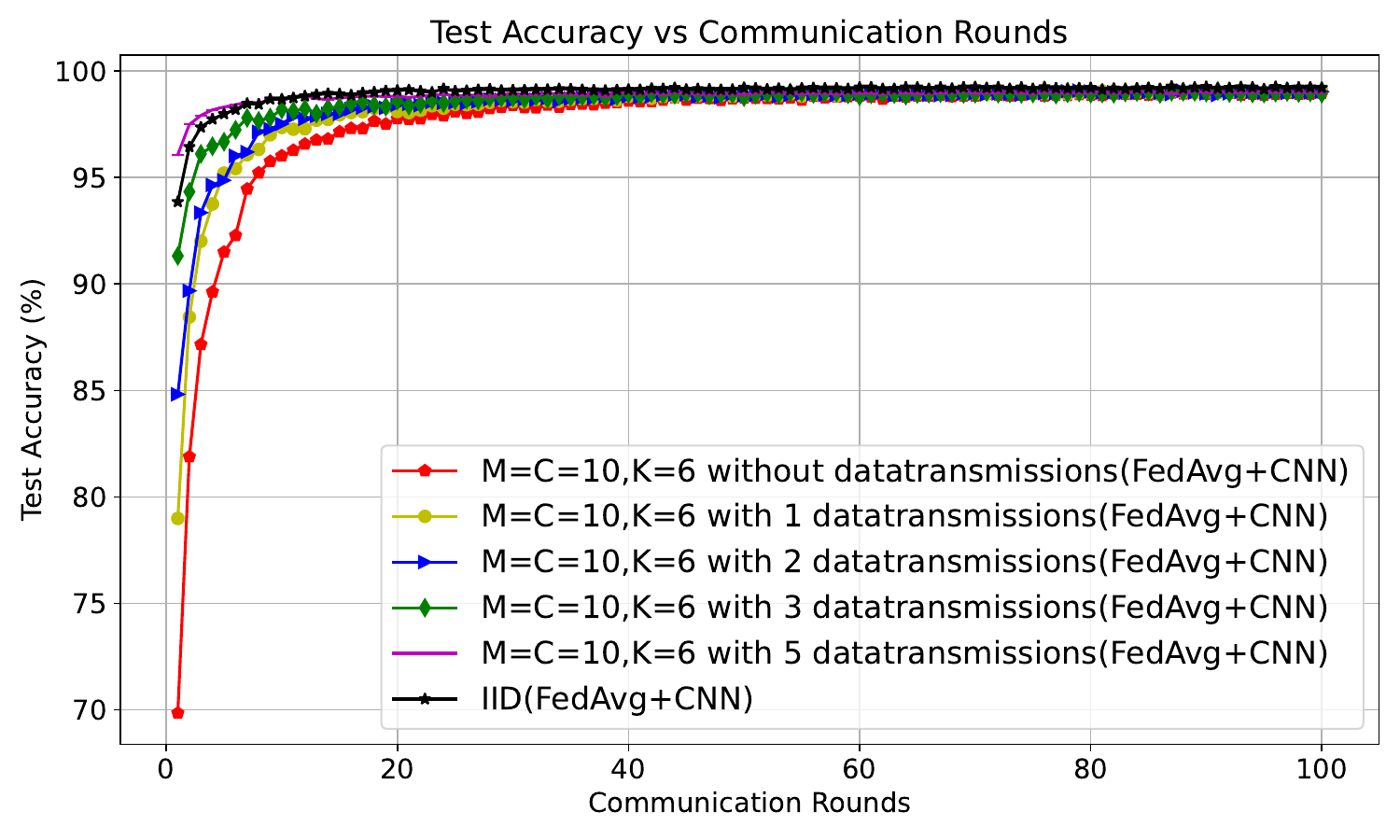} %
\label{fig_3}}
\hfil
\subfloat[]{\includegraphics[width=3.2in]{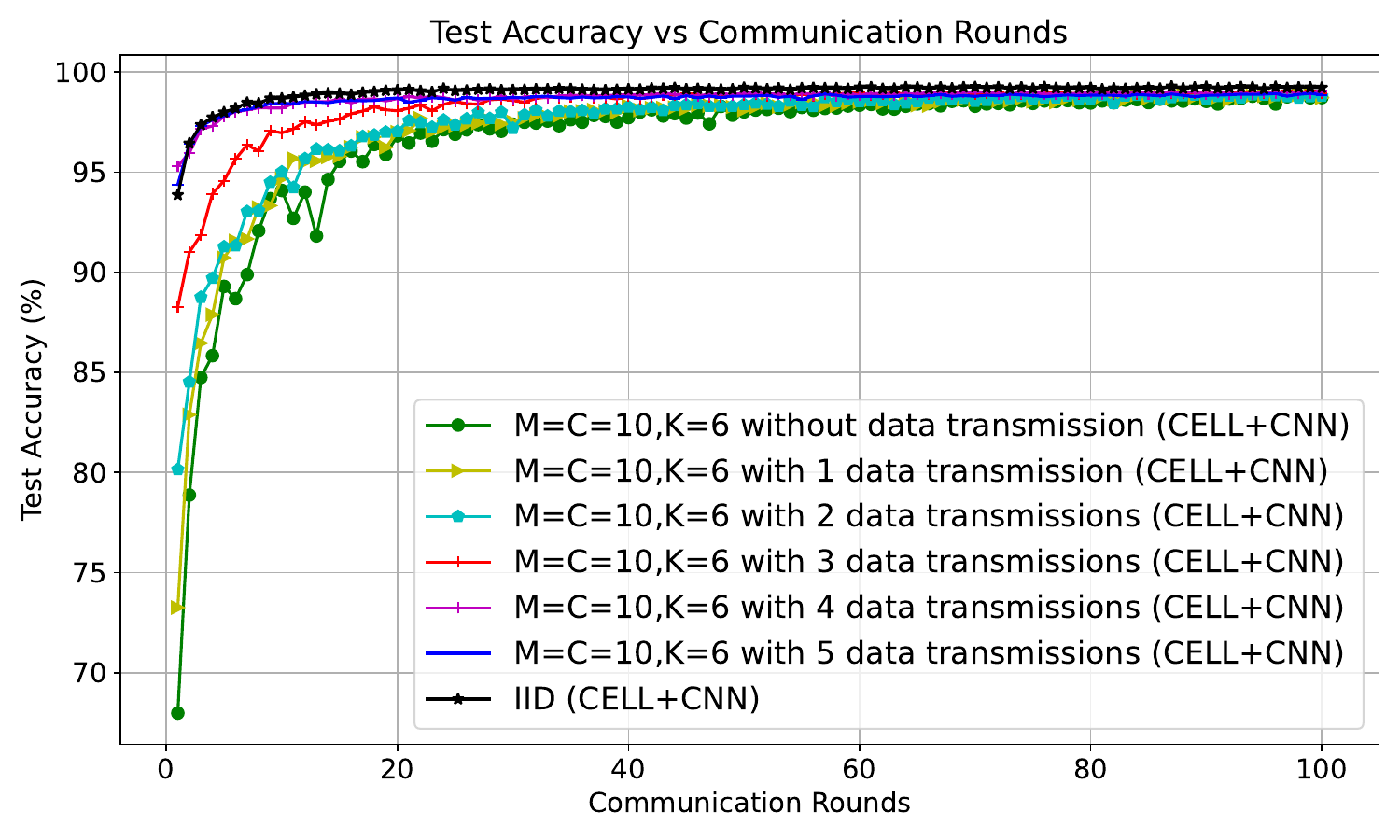}%
\label{fig_4}}
\caption{Accuracy improvement of FL scheme of  $M=C=10$, and $K=6$ when data shuffling prior to FL with CNN is designed by CDPIC($S$,$K$) (a) FL with FedAvg, (b) FL with CELL}
\label{Fig_M10C10K6_3}
\end{figure*}

\begin{figure*}[!t]
\centering
\subfloat[]{\includegraphics[width=3.2in]{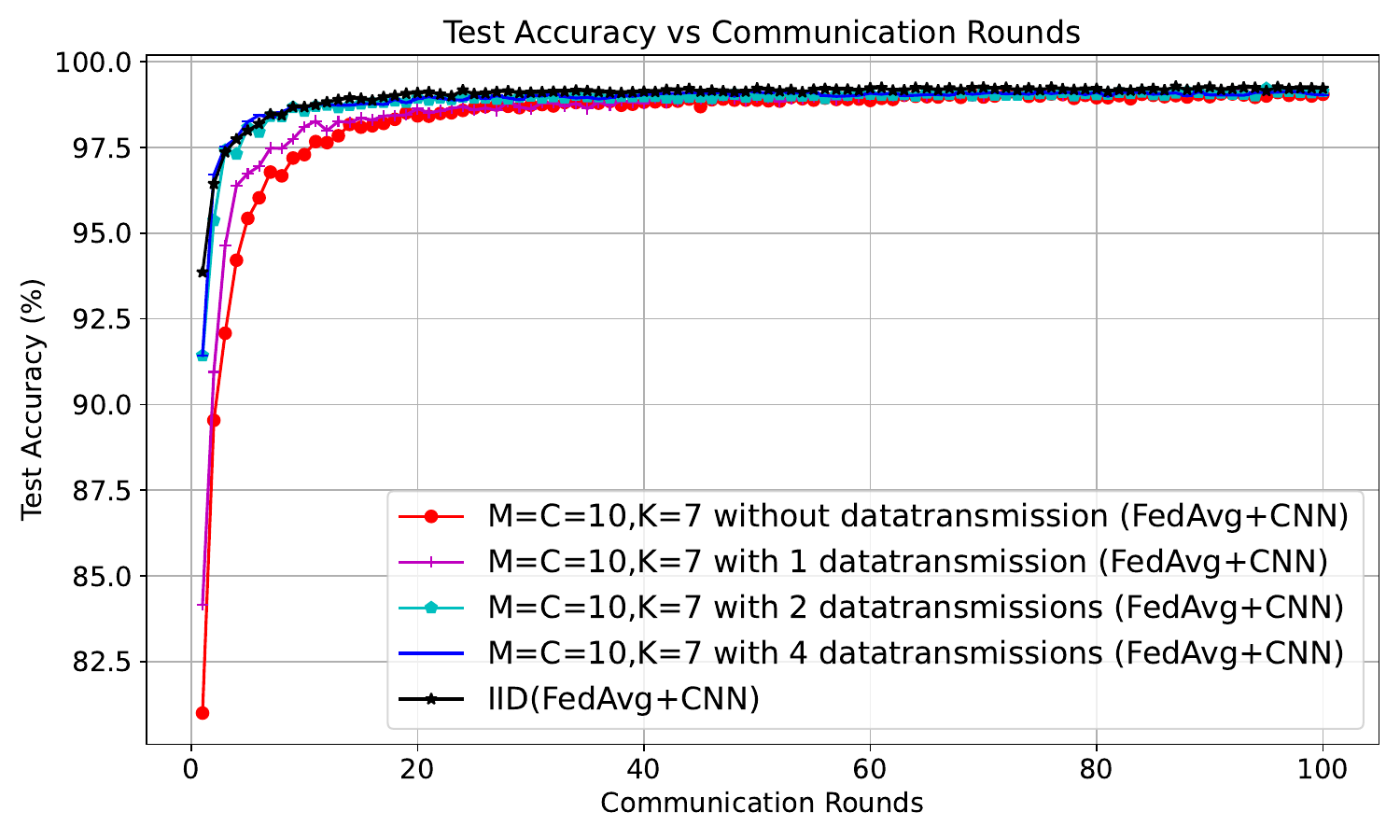}%
\label{fig_first_case}}
\hfil
\subfloat[]{\includegraphics[width=3.2in]{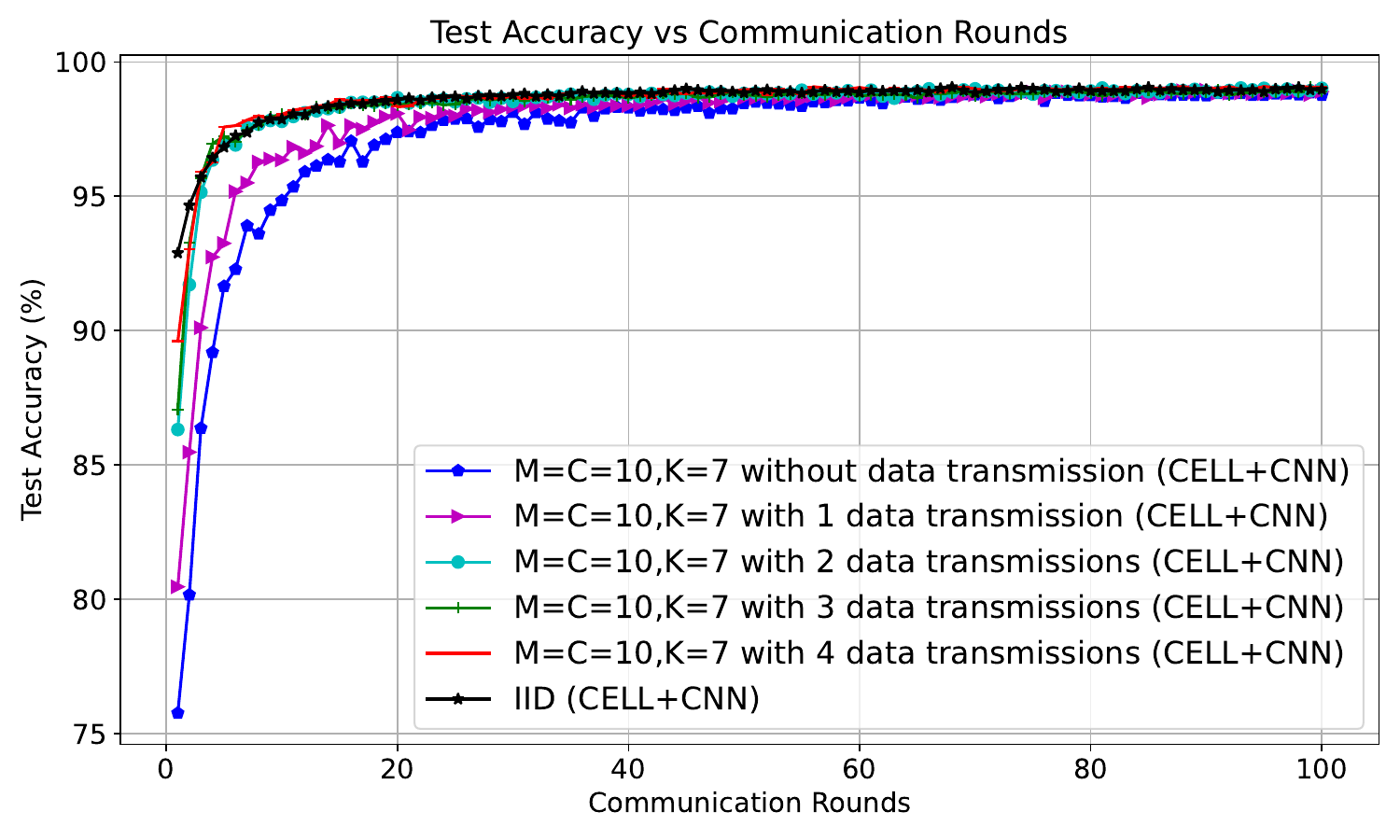}%
\label{fig_second_case}}
\caption{Accuracy improvement of FL scheme of  $M=C=10$, and $K=7$ when data shuffling prior to FL with CNN is designed by CDPIC($S$,$K$) (a) FL with FedAvg, (b) FL with CELL}
\label{Fig_M10C10K7_3}
\end{figure*}

For the case with $M = C = 10$, different CDPIC($S$,$K$) schemes developed in this work for different $K$ and $S$ values are applied. The results with CNN on MNIST data are tabulated in  Table~\ref{Tab:Summary_simulation} and Table \ref{Tab:Summary}. From Table~\ref{Tab:Summary_simulation}, we can see that with $K = 6$, without any data shuffling ($S = 0$), the FL model acquires accuracy around 91 $\%$ within 5 rounds. When $S=1$, each node is required to acquire one additional data class beyond its original $K=6$ classes. Without the CDPIC scheme, this would typically require $N_W = 3 $ transmissions. In contrast, the CDPIC-based data shuffling method explained in Section \ref{K_eqM2} reduces the number of transmissions to $N = 2$ transmissions. This reduction in communication overload is expressed in terms of transmission efficiency as $\frac{(N_W - N)}{N_W} \times 100$, in Table~\ref {Tab:Summary_simulation}. Corresponding to $K=6$ and $S= 1$ transmission efficiency is approximately 33 $\%$. At this point, the FL model achieves around 95$\%$ accuracy. As we further increase $S$, the FL accuracy continues to improve. For example, when $S=2$, the model reaches around 96.5 $\%$ accuracy, and the transmission efficiency improves to 40 $\%$. This trend indicates that each additional transmission helps to reduce the non-IID nature of data and enhances FL performance. To reach an IID-like FL performance, each node may need access to $S=3$ new data classes. Without CDPIC, this would require 10 transmissions, but CDPIC brings this down to 5 transmissions, demonstrating a significant communication saving and improved transmission efficiency as $S$ increases. The corresponding improvement in FL accuracy achieved through CDPIC-based data shuffling for $M=C=10,K=6$ is illustrated in Fig.~\ref{Fig_M10C10K7_3}. Assuming each transmission carries 2000 sample images of size $28 \times 28 $ compressed using JPEG at 90$\%$ quality (allows 80 $\%$ compression for MNIST data), 50 $\%$ transmission efficiency corresponds to saving of 1.2 Mbits ($0.5 \times 2000 \times 28 \times 28 \times 8 \times 0.2 $bits) which is a huge saving in transmission cost. With a transmission rate of 1 Mbps, this corresponds to a reduction in transmission time by 1200 msec.  The results in Table ~\ref{Tab:Summary_simulation} for different  $ M=C,  K \geq \frac{M}{2} + 1$ configurations clearly demonstrate the communication benefits of our approach. The improvement in FL performance corresponding to $ K= 7$ is demonstrated in Fig.~\ref{Fig_M10C10K7_3}.

\begin{table*}[h]
\centering
\caption{Transmission Efficiency in FL using LW-CNN with the Proposed CDPIC($S$,$K$) Schemes }
\label{Tab:Summary_simulation_LWCNN}
\begin{adjustbox}{width=0.9\linewidth}
\renewcommand{\arraystretch}{1.3}
\begin{tabular}{|c|c|c|c|c|c|c|}
\hline
Case          & \multicolumn{1}{c|}{\begin{tabular}[c]{@{}c@{}}Number of  \\ required \\ new data\\  classes\end{tabular}} & \multicolumn{1}{c|}{\begin{tabular}[c]{@{}c@{}}Number of \\ transmissions \\ without using \\ CDPIC Scheme ($N_W$)\end{tabular}} & \multicolumn{1}{c|}{\begin{tabular}[c]{@{}c@{}}Number of \\ transmissions \\ with using CDPIC\\  Scheme ($N$)\end{tabular}} & \multicolumn{1}{c|}{\begin{tabular}[c]{@{}c@{}}Transmission\\  Efficiency\\ ($\%$)     \\  $\frac{(N_W-N)}{N_W} \times 100$\end{tabular}} & \multicolumn{2}{c|}{\begin{tabular}[c]{@{}c@{}}Accuracy improvment \\ after 10 rounds \\($\%$)\end{tabular}} \\  \cline{6-7}
\multicolumn{1}{|c|}{} &     &  & & & FedAvg+LW-CNN & CELL+LW-CNN     \\ \hline

\multirow{5}{*}{\begin{tabular}[c]{@{}c@{}}$M=C=10,K=6$ \\ Sec. \ref{K_eqM2}         \\ Fig. \ref{Fig_M10C10K6_4}\end{tabular}} & $S=0$      & -  &   - &       -    &  87.50& 84.00 \\ \cline{2-7} 
   & $S=1$ & 3         &   2        &   33.33   & 92.50    & 89.00 \\ \cline{2-7} 
           & $S=2$ &    5      &     3      &     40 &   94.00   & 93.00 \\ \cline{2-7} 
           & $S=3$ &   8       &    4       &    50  &   96.80 & 96.50 \\ \cline{2-7} 
           & $S=4$ &  10        &  5         &   50   &  97.50  & 97.40 \\ \hline
\multirow{5}{*}{\begin{tabular}[c]{@{}c@{}}$M=C=10,K=7$\\ Sec. \ref{K_grM2}\\       Fig. \ref{Fig_M10C10K7_4}\end{tabular}}     & $S=0$ &   -       &         -  &   -   &   87.50 & 86.00\\ \cline{2-7} 
           & $S=1$ &  4        &       2    &    50  &   92.00  &92.00\\ \cline{2-7} 
           & $S=2$ &     7     &    3       &     57.14 &   94.80   & 94.00\\ \cline{2-7} 
           & $S=3$ &    10      &     4      &  60    &   97.50    &97.40\\ \hline

\end{tabular}
\end{adjustbox}
\end{table*}
\begin{table*}[h]
\centering
\caption{Reduction in Number of Communication Rounds in FL using LW-CNN with Proposed CDPIC($S$,$K$) Schemes. }
\label{Tab:Summary_LWCNN}
\begin{adjustbox}{width=0.8\linewidth}
\renewcommand{\arraystretch}{1.3}

\begin{tabular}{|l|c|c|c|c|}
\hline

\multicolumn{1}{|c|}{\multirow{2}{*}{Case}} & \multicolumn{1}{c|}{\begin{tabular}[c]{@{}c@{}}Number of  \\ required new data \\ classes to get \\CML accuracy\end{tabular}}& \multicolumn{3}{c|}{\begin{tabular}[c]{@{}l@{}}\\Number of FL rounds  to reach CML Accuracy \\\end{tabular}}            \\ \cline{3-5}
\multicolumn{1}{|c|}{} &       &  & FedAvg+LW-CNN & CELL+LW-CNN     \\ \hline

\multirow{2}{*}{\begin{tabular}[c]{@{}c@{}}$M = C = 10, K = 6  $           \\  Fig. \ref{Fig_M10C10K6_4}\end{tabular}}        & \multirow{2}{*}{\begin{tabular}[c]{@{}c@{}} $S=4$\end{tabular}} &  \multicolumn{1}{c|}{Initial non-IID Data:}     & \multicolumn{1}{c|}{200}       &  300  \\ \cline{3-5}
 &                            & \multicolumn{1}{c|}{After $S+1$   Transmissions:} & \multicolumn{1}{c|}{20}       &  20 \\ \cline{3-5}

 \hline
\multirow{2}{*}{\begin{tabular}[c]{@{}c@{}}$M = C = 10, K = 7$             \\Fig. \ref{Fig_M10C10K7_4}\end{tabular}} & \multirow{2}{*}{\begin{tabular}[c]{@{}c@{}} $S=3$\end{tabular}}  &  \multicolumn{1}{c|}{Initial non-IID Data:}     & \multicolumn{1}{c|}{160}       & 200  \\ \cline{3-5}
&                            & \multicolumn{1}{c|}{After $S+1$ Transmissions:} & \multicolumn{1}{c|}{20}       & 20    
\\ \cline{3-5}

\hline

\end{tabular}

\end{adjustbox}
\end{table*}

\begin{figure}[!t]
    \centering
    \includegraphics[width=3.2in]{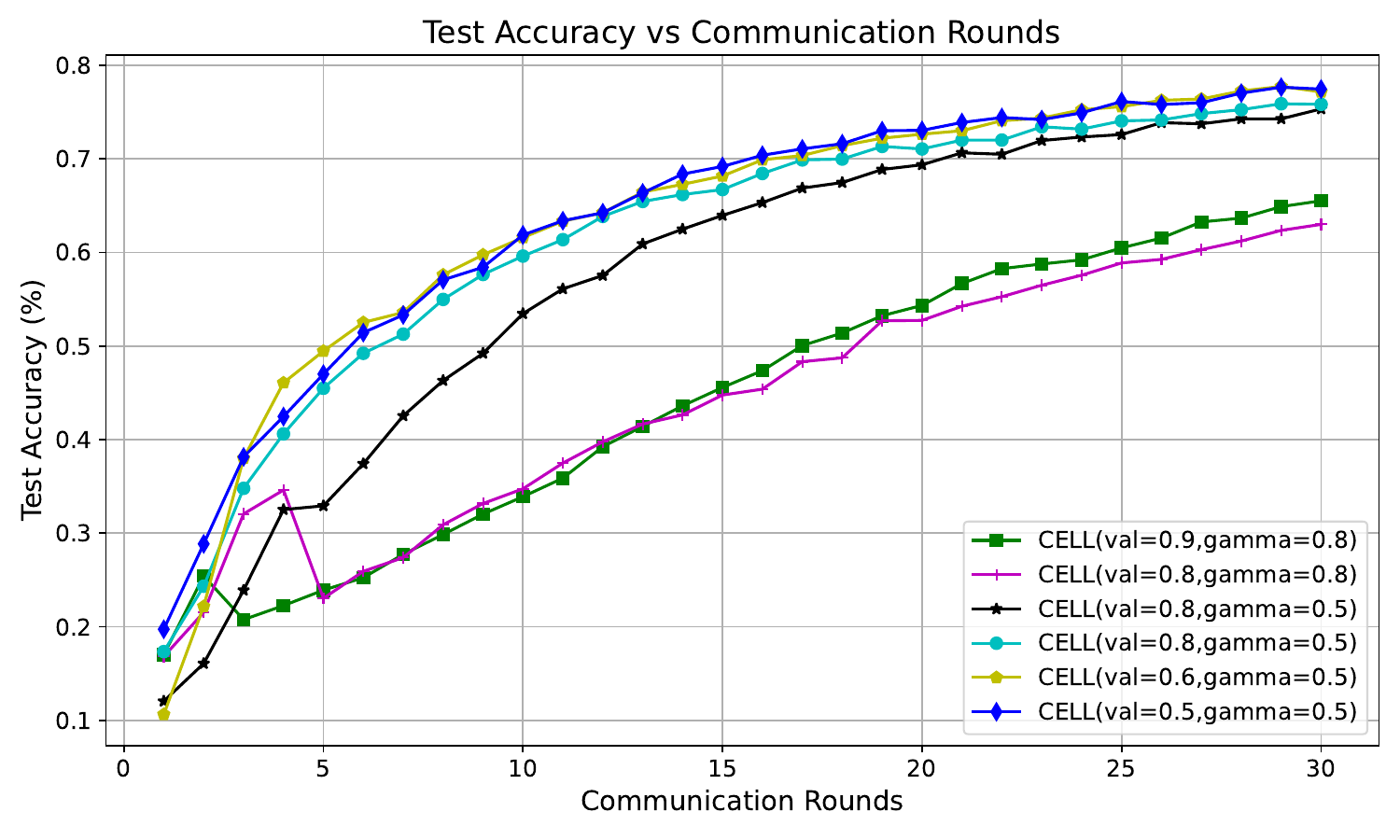}
    \caption{CELL performance under non-IID data distribution for varying validation thresholds (val) and target pruning rates (gamma)}
    \label{fig_6}
\end{figure}

\begin{figure}[!t]
    \centering
    \includegraphics[width=3.2in]{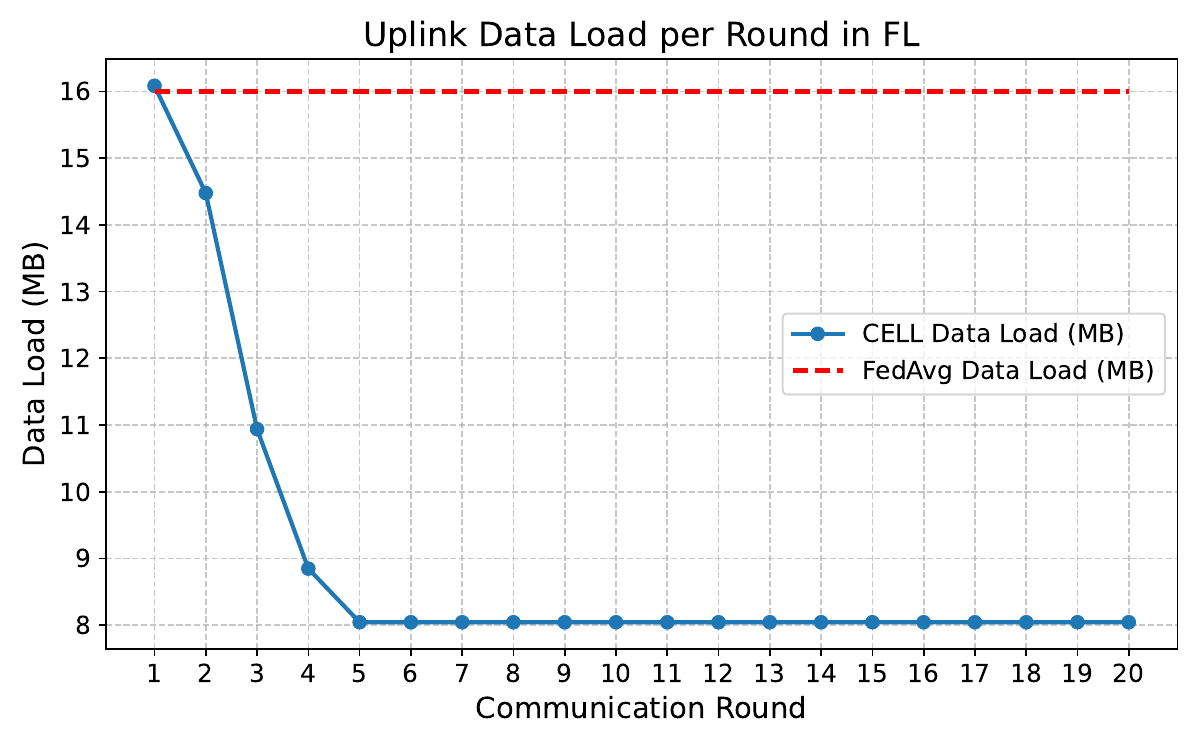}
    \caption{Comparison of Uplink Data Load per Round in FL with CELL vs. FL with FedAvg }
    \label{fig_7}
\end{figure}

Results presented in Table \ref{Tab:Summary} show the reduction in the number of communication rounds of the FL process due to data shuffling, which can bring in a reduction in both the communication overhead and latency. It can be seen that the number of communication rounds reduces significantly for both FedAvg and CELL-based FL schemes for different values of $C$, $K$, and $S$. From the results, it can be seen that CELL takes more rounds to converge than FedAvg, especially when there is more heterogeneity in data distribution. But it must be noted that the number of parameters transmitted between client nodes and CS in a round is very low in CELL compared to FedAvg. This makes CELL more attractive in ITS scenarios where there are tight bandwidth constraints. Specifically, we leveraged CELL to optimize performance by fine-tuning the validation threshold and maximum pruning rate, and the impact of these optimizations in our FL setup is illustrated in Fig.~\ref{fig_6}. The comparison of the number of parameters transmitted in a round for CELL and FedAvg schemes is presented in Fig.~\ref{fig_7}. The results demonstrate that CELL can achieve nearly the same accuracy as FedAvg with much lower transmission cost. For example, from Table \ref{Tab:Summary}, considering the case with $M=C=10$, $K=6$, $S=3$, FedAvg requires 10 communication rounds while CELL requires 13 rounds for CML accuracy. From Fig.~\ref{fig_7}, it can be seen that the uplink cost per transmission is 16 MB for FedAvg while it reduces to 8 MB after the fifth round for CELL. This means the total transmission cost of FedAvg is 160 MB while it is 124 MB for CELL.

\begin{figure*}[!t]
\centering
\subfloat[]{\includegraphics[width=3.2in]{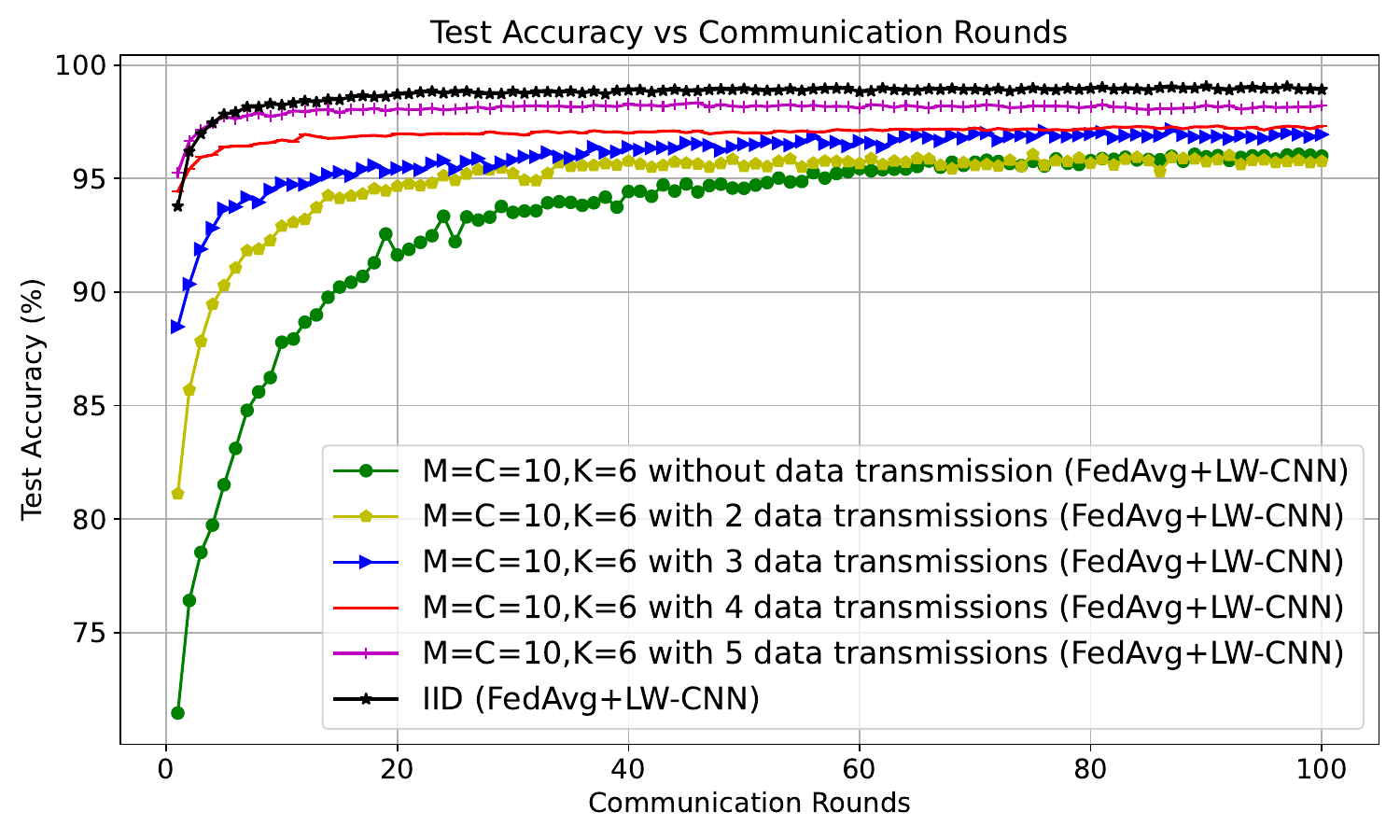}%
\label{fig_first_case}}
\hfil
\subfloat[]{\includegraphics[width=3.2in]{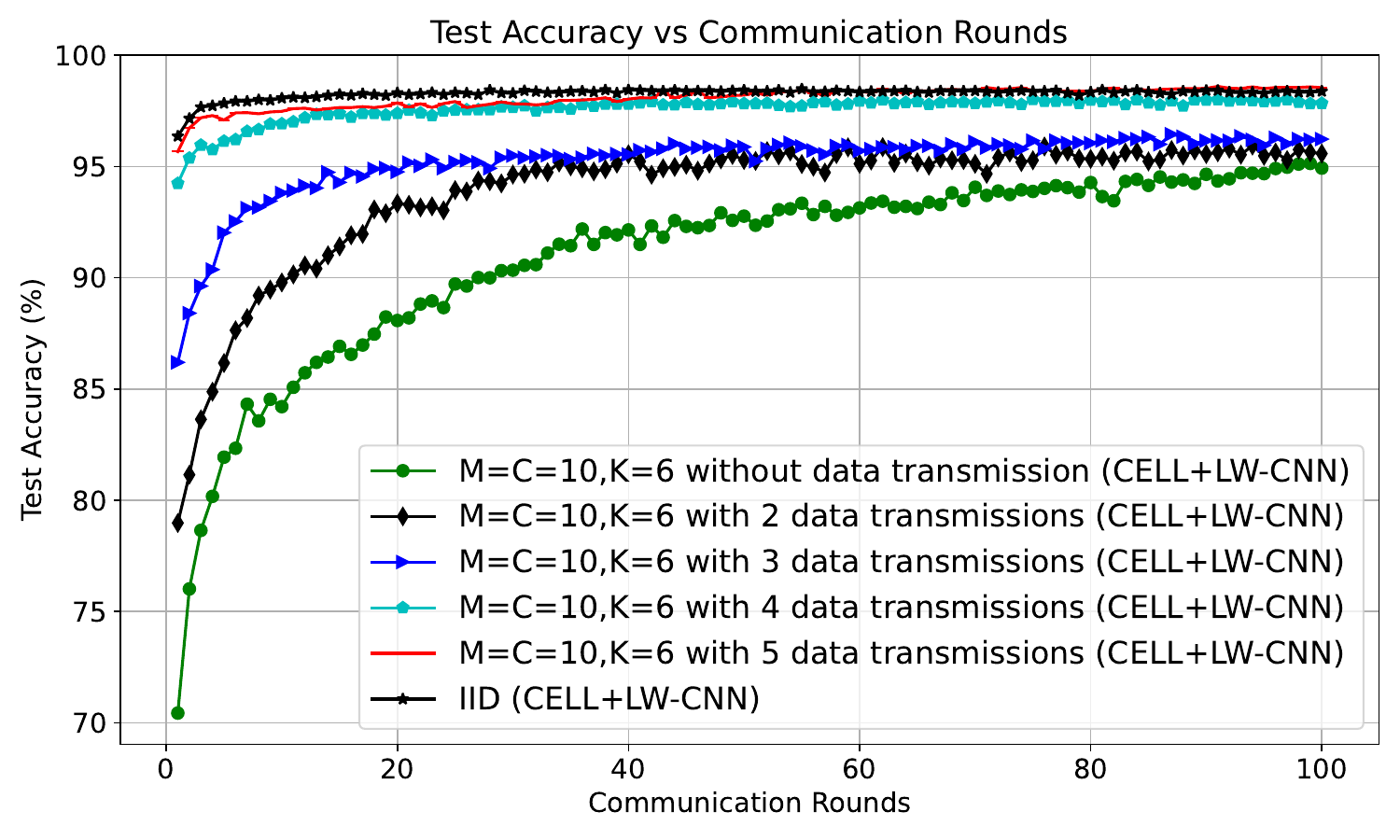}%
\label{fig_second_case}}
\caption{Accuracy improvement of FL scheme of  $M=C=10$, and $K=6$ when data shuffling prior to FL with LW-CNN is designed by CDPIC($S$,$K$) (a) FL with FedAvg, (b) FL with CELL}
\label{Fig_M10C10K6_4}
\end{figure*}

\begin{figure*}[!t]
\centering
\subfloat[]{\includegraphics[width=3.2in]{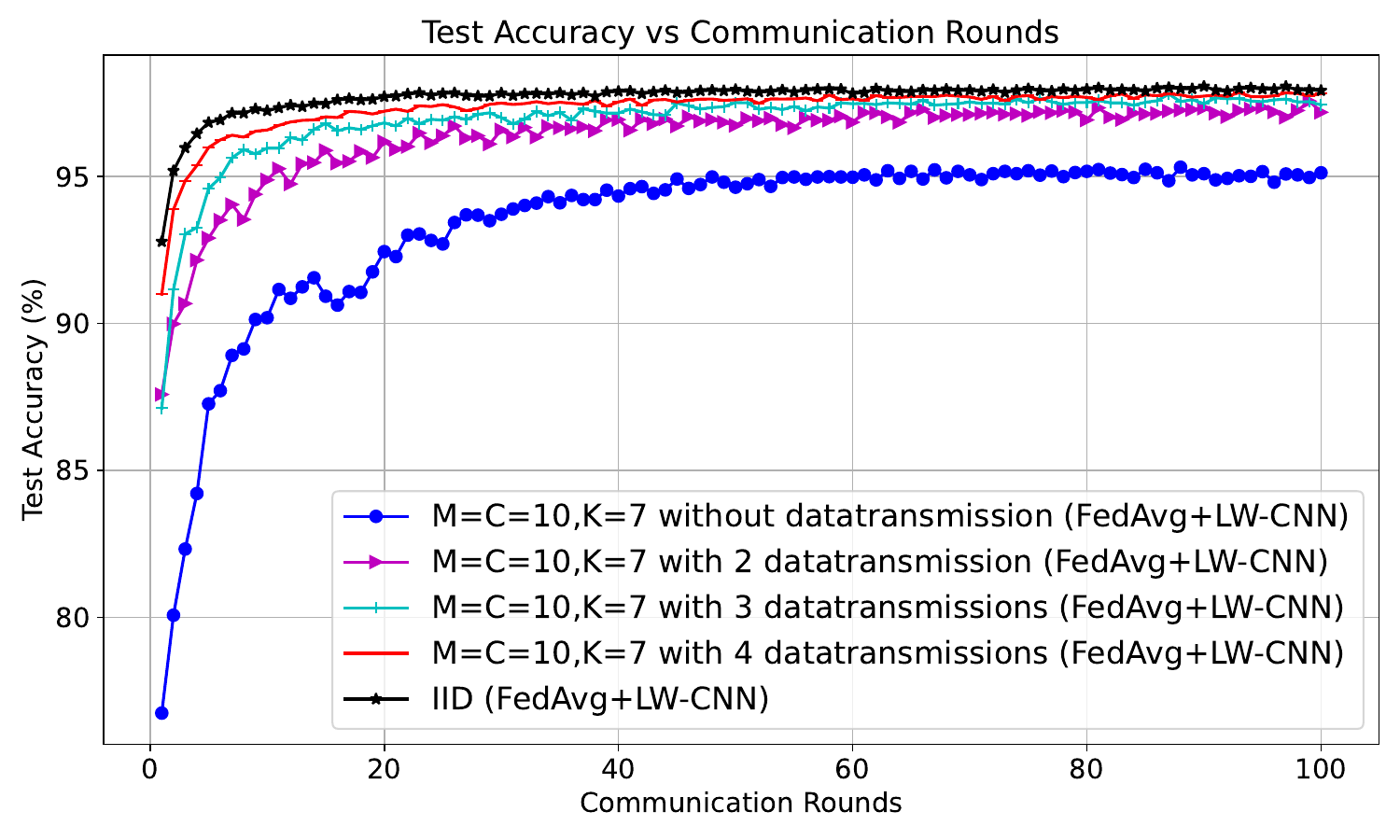}
\label{fig_first_case}}
\hfil
\subfloat[]{\includegraphics[width=3.2in]{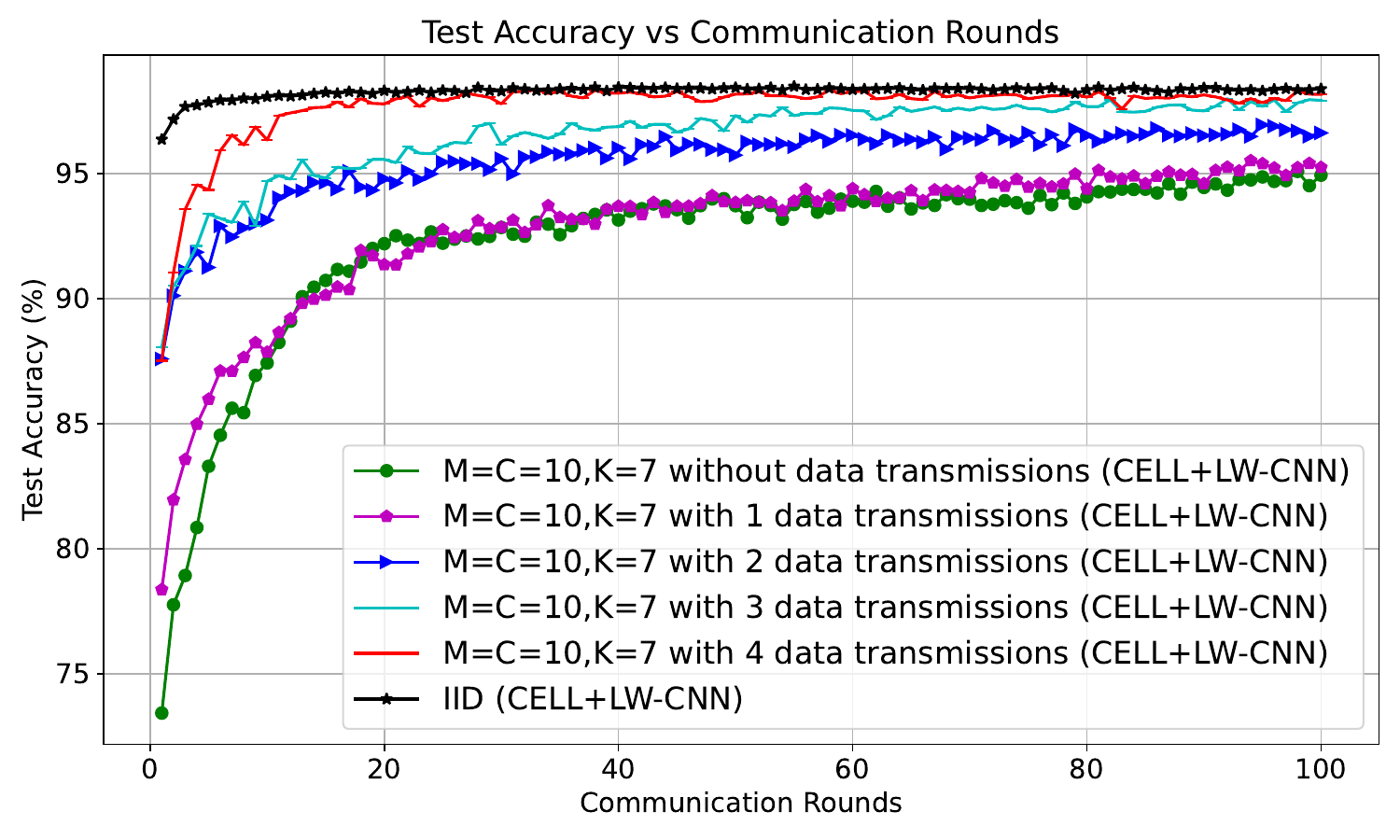}%
\label{fig_second_case}}
\caption{Accuracy improvement of FL scheme of  $M=C=10$ and $K=7$ when data shuffling prior to FL with LW-CNN is designed by CDPIC($S$,$K$) (a) FL with FedAvg, (b) FL with CELL}
\label{Fig_M10C10K7_4}
\end{figure*}

All the experiments are repeated by replacing CNN with the LW-CNN network, and similar results are obtained (Fig. \ref{Fig_M10C10K6_4}-Fig. \ref{Fig_M10C10K7_4}). The CML accuracy with LW-CNN is approximately 97.5 $\%$. Under IID conditions, FL with LW-CNN converges to CML accuracy within approximately 10 communication rounds, slower compared to CNN, which converges in 5 rounds. Therefore, to ensure a consistent and fair comparison across all data shuffling levels, FL performance with LW-CNN is evaluated at the 10th round. The corresponding experimental results are summarized in Table \ref{Tab:Summary_simulation_LWCNN} and Table \ref{Tab:Summary_LWCNN}. The results show that LW-CNN considerably lowers transmission overhead when compared to the general CNN model. However, LW-CNN needs more communication rounds to reach the same target accuracy, which results in slower convergence.


\begin{table*}[!t]
\centering
\caption{Transmission Efficiency in FL using CIFAR10 with the Proposed CDPIC($S$,$K$) Schemes }
\label{Tab:Summary_simulation_CIFAR10}
\begin{adjustbox}{width=0.9\linewidth}
\renewcommand{\arraystretch}{1.3}
\begin{tabular}{|c|c|c|c|c|c|}
\hline
Case          & \multicolumn{1}{c|}{\begin{tabular}[c]{@{}c@{}}Number of  \\ required \\ new data\\  classes\end{tabular}} & \multicolumn{1}{c|}{\begin{tabular}[c]{@{}c@{}}Number of \\ transmissions \\ without using \\ CDPIC Scheme ($N_W$)\end{tabular}} & \multicolumn{1}{c|}{\begin{tabular}[c]{@{}c@{}}Number of \\ transmissions \\ with using CDPIC\\  Scheme ($N$)\end{tabular}} & \multicolumn{1}{c|}{\begin{tabular}[c]{@{}c@{}}Transmission\\  Efficiency\\  $\frac{(N_W-N)}{N_W} \times 100$\\ ($\%$)     \end{tabular}} & \multicolumn{1}{c|}{\begin{tabular}[c]{@{}c@{}}Accuracy improvment \\ after 20 rounds \\($\%$)\end{tabular}} \\ \hline

\multirow{5}{*}{\begin{tabular}[c]{@{}c@{}}$M=C=10,K=6$          \\ Fig. \ref{fig_CIFAR}\end{tabular}} & $S=0$      & -  &   - &       -    &  45.00\\ \cline{2-6} 
   & $S=1$ & 3         &   2        &   33.33   & 55.00     \\ \cline{2-6} 
           & $S=2$ &    5      &     3      &     40 &  60.00  \\ \cline{2-6} 
           & $S=3$ &   8       &    4       &    50  &   64.00  \\ \cline{2-6} 
           & $S=4$ &  10        &  5         &   50   &  68.00   \\ \hline

\end{tabular}
\end{adjustbox}
\end{table*}

\begin{figure}[!t]
    \centering
    \includegraphics[width=3.2in]{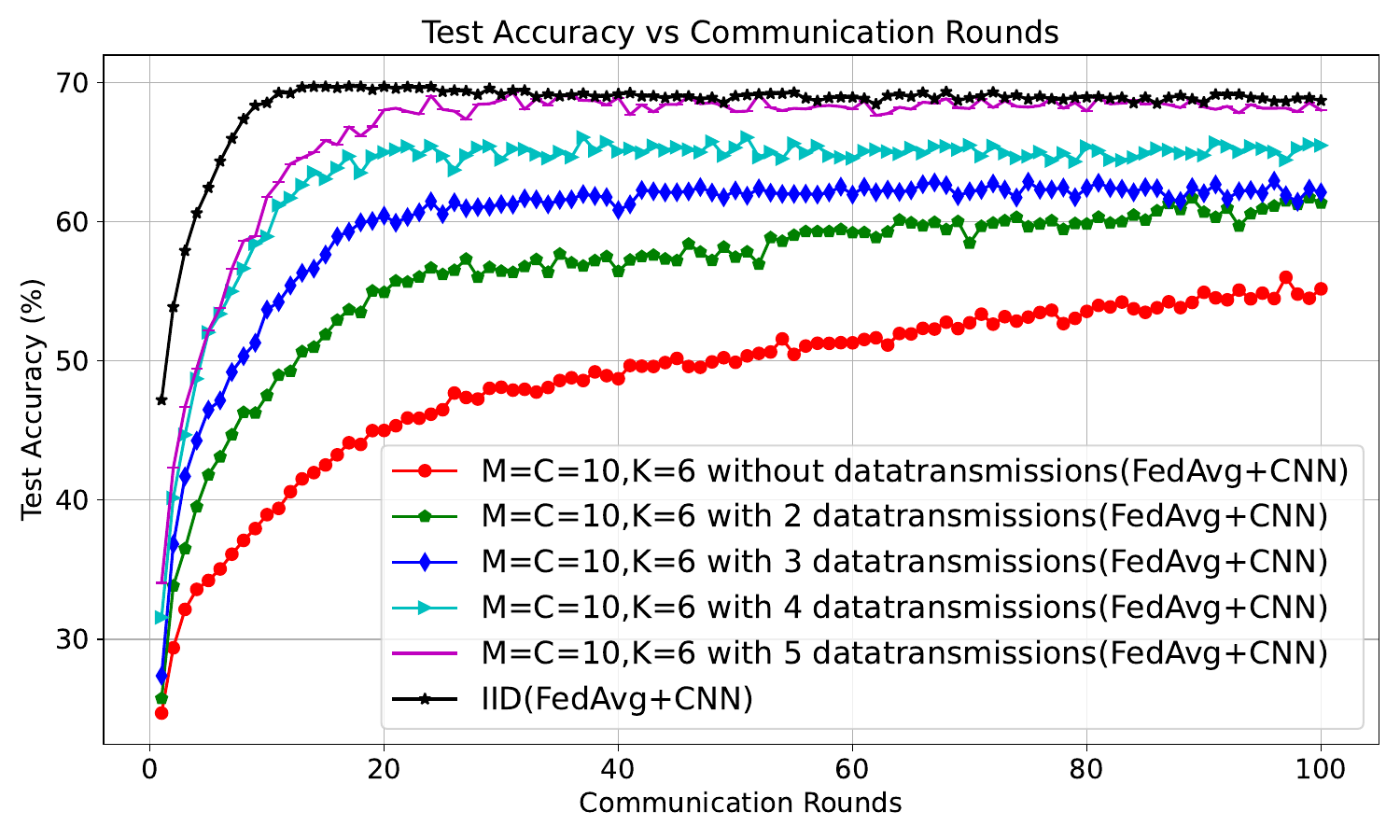}
    \caption{Accuracy improvement of FL scheme on the CIFAR-10 dataset with $M=C=10$ and $K=6$ when data shuffling prior to FL with CNN is designed by CDPIC($S$,$K$) }
    \label{fig_CIFAR}
\end{figure}

To further validate the effectiveness and generalizability of the proposed CDPIC($S$,$K$)-based data shuffling scheme, we extended our experiments to the CIFAR-10 dataset under the same
configuration of $M = C = 10$, $K = 6$. The corresponding experimental results are presented in Fig. \ref{fig_CIFAR} and summarized in Table \ref{Tab:Summary_simulation_CIFAR10}. In this case, without any data transmissions ($S = 0$), the FL accuracy saturates at around 55 $\%$. However, after just 5 CDPIC transmissions, the model performance improves significantly, achieving accuracy levels comparable to that of the IID setup. These results reinforce the capability of the proposed CDPIC scheme to enhance transmission efficiency in the data shuffling phase while transforming non-IID FL scenarios to approximate IID conditions.

In summary, the advantage due to improved communication efficiency achieved by applying the proposed CDPIC schemes in an FL scenario is twofold. (i) Reduction in transmission energy and (ii) Reduction in latency during the FL process. Reduction in the number of transmissions brings in significant reduction in transmission time. It is shown that even with a very simple dataset such as MNIST, a reduction in transmission time of the order of 1200 msec can be achieved. Since each coded transmission will benefit more nodes in FL, the convergence speed also improves. This reduction in latency in multiple phases is an important advantage for latency-sensitive ITS applications. The performance improvement achieved by the proposed CDPIC($S$,$K$) schemes is validated with different datasets (MNIST, CIFAR10), different network architectures (CNN, LW-CNN), and different FL processes (FedAvg, CELL).

\section{Conclusion}
\label{sec:conclusion}
In this work, we consider a Federated Learning application in Intelligent Transportation Systems. Here, a VANET scenario is considered where RSUs are the worker nodes in FL. Considering a circular/ rectangular area around a central server, individual RSU nodes can have non-IID data distribution with overlap among data sets of different RSUs. The experimental results show that there is a significant reduction in the FL convergence performance due to non-IID data distribution. Further, this work proposes the design of suitable CDPIC schemes to implement data shuffling with excellent transmission efficiency. We established a lower bound for the general decentralized pliable index coding problem DPIC($S$), demonstrating that at least $S+1$ transmissions are required to satisfy client demands. FL experiments with data shuffling using CDPIC($S$,$K$) schemes clearly show an improvement in communication overhead and latency through a reduction in the number of communication rounds in the FL process. To the best of our knowledge, this is the first work that applies pliable index coding solutions to improve the performance of FL and FSL. Experiments are carried out with the popular FedAvg FL system to validate the improvement in communication efficiency. Considering delay-sensitive ITS applications with very tight bandwidth constraints, experiments are also conducted with a futuristic federated sub-model learning technique called CELL. The improved performance is validated using both CNN and LW-CNN network architectures. These results show that our proposed approach generalizes well across different model architectures as well as different FL techniques, ensuring effective communication-efficient FL for ITS applications.

\section{Acknowledgement }
The authors gratefully acknowledge the Centre for Computational Modelling and Simulation (CCMS) and the Central Computer Centre (CCC) at NIT Calicut for providing access to the NVIDIA DGX Station facility. They also express their sincere gratitude to the Department of Science and Technology (DST), Government of India, for supporting this work under the FIST scheme No. $SR/FST/ET-I/2017/68$. Partial support for this work was provided through the J.C. Bose National Fellowship awarded to Prof. B. Sundar Rajan by the Science and Engineering Research Board (SERB), DST, Government of India, as well as through the SERB Startup Research Grant and the DST INSPIRE Faculty Fellowship awarded to Dr. Nujoom Sageer Karat.

\bibliographystyle{IEEEtran}
\bibliography{reference}



\end{document}